\documentclass[manuscript]{aastex6}

\usepackage{xspace}
\usepackage{capt-of}   
\usepackage{graphicx}
\usepackage{numprint}
\npdecimalsign{.}   
\nprounddigits{3}

\usepackage{lineno}

\newcommand{\pound}{%
  {\settoheight{\dimen0}{C}\kern-.05em \resizebox{!}{\dimen0}{\raisebox{\depth}{\#}}}}

\newcommand{\twopr}{^{\prime \prime}}

\newcommand{\ed}{\textcolor{black}}
\newcommand{\edd}{\textcolor{black}}

\newcommand\rf{\textcolor{black}}

\newcommand{\sncu}{SN~2012cu\xspace}
\newcommand{\snfe}{SN~2011fe\xspace}
\newcommand{\AV}{$A_V$\xspace}
\newcommand{\RV}{$R_V$\xspace}
\newcommand{\EBV}{$E(B-V)$\xspace}
\newcommand{\Delmu}{$\Delta \mu$\xspace}
\newcommand{\chisqnu}{$\chi^2_\nu$\xspace}
\newcommand{\EWDIB}{EW$_\mathrm{DIB}$(5780 \AA)\xspace}

\newcommand{\EBVI}{0.983}
\newcommand{\eEBVI}{0.046}
\newcommand{\RVI}{3.035}
\newcommand{\eRVI}{0.131}

\newcommand{\AVI}{2.973}
\newcommand{\eAVI}{0.051}
\newcommand{\DelmuI}{1.952}
\newcommand{\eDelmuI}{0.068}

\newcommand{\EBVII}{1.002}
\newcommand{\eEBVII}{0.031}
\newcommand{\RVII}{2.952}
\newcommand{\eRVII}{0.081}

\newcommand{\AVII}{2.944}
\newcommand{\eAVII}{0.043}
\newcommand{\DelmuII}{1.982}
\newcommand{\eDelmuII}{0.061}

\newcommand{\EBVIII}{0.999}
\newcommand{\eEBVIII}{0.029}
\newcommand{\RVIII}{2.963}
\newcommand{\eRVIII}{0.079}

\newcommand{\AVIII}{2.950}
\newcommand{\eAVIII}{0.048}
\newcommand{\DelmuIII}{1.972}
\newcommand{\eDelmuIII}{0.067}

\newcommand{\cuPzero}{$-6.8$\xspace}
\newcommand{\cuPone}{$-3.8$\xspace}
\newcommand{\cuPtwo}{$-1.7$\xspace}
\newcommand{\cuPthree}{3.3\xspace}
\newcommand{\cuPfour}{6.3\xspace}

\newcommand{\cuPsix}{11.2\xspace}

\newcommand{\cuPeleven}{23.2\xspace}
\newcommand{\cuPsixteen}{46.2\xspace}

\newcommand{\fePtwo}{$-2.0$\xspace}

\newcommand{\fePsix}{11.0\xspace}

\newcommand{\Cite}[2]{(for a review on NGC 5128, see~\citealt{#1}; for NGC 4772, see~\citealt{#2})}

\begin{document}


\title{The Extinction properties of and distance to the highly reddened Type~Ia supernova SN 2012cu}



\author
{
    X.~Huang,\altaffilmark{1}
    Z.~Raha,\altaffilmark{1}
    G.~Aldering,\altaffilmark{2}
    P.~Antilogus,\altaffilmark{4}
    S.~Bailey,\altaffilmark{2}
    C.~Baltay,\altaffilmark{5}
    K.~Barbary,\altaffilmark{3}
    D.~Baugh,\altaffilmark{6}
    K.~Boone,\altaffilmark{2,3}
    S.~Bongard,\altaffilmark{4}
    C.~Buton,\altaffilmark{7}
    J.~Chen,\altaffilmark{6}
    N.~Chotard,\altaffilmark{7}
    Y.~Copin,\altaffilmark{7}
    P.~Fagrelius,\altaffilmark{2,3}
    H.~K.~Fakhouri,\altaffilmark{2,3}
    U.~Feindt,\altaffilmark{16}
    D.~Fouchez,\altaffilmark{9}
    E.~Gangler,\altaffilmark{10}  
    B.~Hayden,\altaffilmark{2}
    W.~Hillebrandt,\altaffilmark{15}
    A.~G.~Kim,\altaffilmark{2}
    M.~Kowalski,\altaffilmark{8,11}
    P.-F.~Leget,\altaffilmark{10}
    S.~Lombardo,\altaffilmark{8}
    J.~Nordin,\altaffilmark{2,8}
    R.~Pain,\altaffilmark{4} 
    E.~Pecontal,\altaffilmark{12}
    R.~Pereira,\altaffilmark{7}
    S.~Perlmutter,\altaffilmark{2,3}
    D.~Rabinowitz,\altaffilmark{5}
    M.~Rigault,\altaffilmark{8} 
    D.~Rubin,\altaffilmark{2,13}
    K.~Runge,\altaffilmark{2}
    C.~Saunders,\altaffilmark{2,3}
    G.~Smadja,\altaffilmark{7} 
    C.~Sofiatti,\altaffilmark{2,3} 
    A.~Stocker,\altaffilmark{17}
    N.~Suzuki,\altaffilmark{2}
    S.~Taubenberger,\altaffilmark{15}
    C.~Tao,\altaffilmark{6,9}
    R.~C.~Thomas \altaffilmark{14} \\
    (The Nearby Supernova Factory)
}

\altaffiltext{1}
{Department of Physics and Astronomy, University of San Francisco, San Francisco, CA 94117-1080}

\altaffiltext{2}
{
    Physics Division, Lawrence Berkeley National Laboratory, 
    1 Cyclotron Road, Berkeley, CA, 94720
}
\altaffiltext{3}
{
    Department of Physics, University of California Berkeley,
    366 LeConte Hall MC 7300, Berkeley, CA, 94720-7300
}
\altaffiltext{4}
{
    Laboratoire de Physique Nucl\'eaire et des Hautes \'Energies,
    Universit\'e Pierre et Marie Curie Paris 6, Universit\'e Paris Diderot Paris 7, CNRS-IN2P3, 
    4 place Jussieu, 75252 Paris Cedex 05, France
}
\altaffiltext{5}
{
    Department of Physics, Yale University, 
    New Haven, CT, 06250-8121
}
\altaffiltext{6}
{
    Tsinghua Center for Astrophysics, Tsinghua University, Beijing 100084, China 
}
\altaffiltext{7}
{
    Universit\'e de Lyon, F-69622, Lyon, France ; Universit\'e de Lyon 1, Villeurbanne ; 
    CNRS/IN2P3, Institut de Physique Nucl\'eaire de Lyon.
}
\altaffiltext{8}
{
    Institut fur Physik,  Humboldt-Universitat zu Berlin,
    Newtonstr. 15, 12489 Berlin
}
\altaffiltext{9}
{
    Centre de Physique des Particules de Marseille, 
    Aix-Marseille Universit\'e , CNRS/IN2P3, 
    163 avenue de Luminy - Case 902 - 13288 Marseille Cedex 09, France
}
\altaffiltext{10}
{
    Clermont Universit\'e, Universit\'e Blaise Pascal, CNRS/IN2P3, Laboratoire de Physique Corpusculaire,
    BP 10448, F-63000 Clermont-Ferrand, France
}
\altaffiltext{11}
{
    DESY, D-15735 Zeuthen, Germany
}
\altaffiltext{12}
{
    Centre de Recherche Astronomique de Lyon, Universit\'e Lyon 1,
    9 Avenue Charles Andr\'e, 69561 Saint Genis Laval Cedex, France
}
\altaffiltext{13}
{
    Space Telescope Science Institute,
    3700 San Martin Drive, Baltimore, MD 21218
}
\altaffiltext{14}
{
    Computational Cosmology Center, Computational Research Division, Lawrence Berkeley National Laboratory, 
    1 Cyclotron Road MS 50B-4206, Berkeley, CA, 94720
}

\altaffiltext{15}
{
    Max-Planck-Institut f\"ur Astrophysik, Karl-Schwarzschild-Str. 1,
D-85748 Garching, Germany
}

\altaffiltext{16}
{
    The Oskar Klein Centre, Department of Physics, AlbaNova, Stockholm University, 
SE-106 91 Stockholm, Sweden
}

\altaffiltext{17}
{
    Department of Mathematics, University of Colorado,
    Campus Box 395, 
    Boulder, Colorado 80309-0395
}

\begin{abstract}
Correction of Type Ia Supernova brightnesses for extinction by dust has proven to be a vexing problem.  Here we study the dust foreground to the highly reddened SN 2012cu, which is projected onto a dust lane in the galaxy NGC 4772.  The analysis is based on multi-epoch, spectrophotometric observations spanning 3,300 - 9,200 \AA,
obtained by the Nearby Supernova Factory.  Phase-matched comparison of the spectroscopically twinned SN 2012cu and SN 2011fe \edd{across 10 epochs} results in the best-fit color excess of \ed{(\EBV, RMS) = (\nprounddigits{2}$\numprint{\EBVII}, \numprint{\eEBVII}$) and total-to-selective extinction ratio of (\RV, RMS) = ($\numprint{\RVII}, \numprint{\eRVII}$) toward SN 2012cu within its host galaxy.}  We further identify several diffuse interstellar bands, and compare the 5780 \AA\ band with the dust-to-band ratio for the Milky Way.  Overall, we find the foreground dust-extinction properties for SN 2012cu to be consistent with those of the Milky Way.  \edd{Furthermore we find no evidence for significant time variation in any of these extinction tracers.}  We also compare the dust extinction curve models of \citet{cardelli89a}, \citet{odonnell94a}, and \citet{fitzpatrick99a}, and find the predictions of \citet{fitzpatrick99a} fit SN~2012cu the best.  Finally, the distance to NGC4772, the host of SN~2012cu, at a redshift of $z = 0.0035$, often assigned to the Virgo Southern Extension, is determined to be 16.6$\pm$1.1 Mpc.  We compare this result with distance measurements in the literature.
\end{abstract}


\section{Introduction} \label{sec:intro}
The \edd{distance measurements} of Type Ia supernovae (SNe Ia) led to the discovery of the accelerated expansion of the universe \citep{riess98a, perlmutter99a}.  This technique, based on the standardizability of the peak luminosity of SNe Ia, remains one of the most powerful probes to understand the cause of the acceleration \citep[e.g.,][]{suzuki12a, betoule14a, rest14a}.  This is especially true in light of recent developments in standardization techniques that promise to substantially reduce brightness dispersion from the canonical value of $\sim0.15$~mag, alternatively based on the incorporation of near-IR light curves, Si II $\lambda$6355 \AA\ absorption line velocity, and local star formation rate \citep{mandel11a, barone-nugent12a, wang09a, foley11a, rigault13a, kelly15a}.  \ed{Recently,} \citet{fakhouri15a} proposed a new standardization procedure based on a spectroscopic twinning method in the optical range that produces a dispersion that is as good as the best of the other standardization methods.  With the prospect that the variation due to the astrophysical differences of SNe Ia can be further \edd{minimized} with the twinning method, dust remains one of the last \edd{key} sources of systematic uncertainty.

Following \citet{cardelli89a}, dust extinction properties are often characterized by one parameter, $R_V \equiv A_V/E(B-V)$, the total-to-selective extinction ratio.   The interstellar dust in the Milky Way (MW) has a well-known average $R_V$ of 3.1 \citep[e.g.,][]{draine03a, schlafly11a}.  The slope of the color standardization for SNe Ia is determined by minimizing the rest-frame $B$-band luminosity ($M_B$) scatter \citep[e.g.][]{tripp98a, tripp99a}.  The slope, usually denoted as $\beta$ \citep[e.g.,][Equation 5]{guy10a}, could naively be expected to equal $R_B$ (= \RV + 1), but likely mixes the effects of SN intrinsic color variation and dust.  In fact, typical fits for $\beta$ yield much lower values \citep[e.g.,][]{tripp98a, hicken09a, wang09a} than expected given the MW average \RV.  

\citet{chotard11a} shed light on this apparent discrepancy.  \ed{Exploiting the fact that spectral features are independent of dust extinction, they first showed that by applying corrections based on the equivalent widths (EW's) of Si II $\lambda$4131 \AA\ and Ca II H\&K features, two highly variable regions in the SN spectra, the mean empirical extinction law obtained by minimizing $M_B$ scatter for their SN sample became much smoother and conformed well to the extinction law of \citet[][CCM89]{cardelli89a}.  Secondly, by adopting a color covariance matrix that is different from the one \edd{based on the output of the SALT2 lightcurve fitter}\citep{guy07a}, they obtained \RV = 2.8$\pm$0.3, in agreement with the average MW value.}
It remains the case, however, that many highly reddened SNe tend to have $R_V$ lower than 2 \citep[e.g.,][]{mandel11a, phillips13a}.  A recent example is SN 2014J, a well-observed, highly reddened SN Ia with a \ed{$R_V$ of 1.4 -- 1.7 and \EBV of 1.2 -- 1.4 \citep{goobar14a, amanullah14a, foley14a, marion15a, brown15a}.
These results are not derived from brightness-color behavior, as with the $\beta$ parameter mentioned above, but rather are obtained from a fit to the shape of the SN spectral energy distribution (SED) using an extinction curve parametrized by \RV.}   

Unlike for cosmological studies, in order to understand the extinction properties of host galaxies, highly reddened SNe are useful.  Comparing a reddened SN Ia with its spectroscopic twin that is little affected by dust would allow the study of host-galaxy extinction without the confusion of intrinsic color variation.  SN 2012cu presents such an opportunity\ed{, as it has one of the highest extinction values observed to date.}  Furthermore, SN 2011fe, in many regards the best studied SN~Ia, is a spectroscopic twin of \sncu.  The fact that \snfe has little or no dust \citep{nugent11a} is also quite important, since comparison to \sncu will provide the total --- not just differential --- extinction.  In this paper, with SN 2011fe as the template, we analyze the wavelength-dependent extinction of SN~2012cu using \ed{a spectrophotometric time series} obtained by the Nearby Supernova Factory \citep[SNfactory,][]{aldering02a}.  \rf{While the nearly ideal pairing between \sncu and \snfe perhaps is rare today, such samples should continue to grow.  The \RV distribution found nearby will then inform high-redshift studies, just as lightcurve parameters now do.}   

With SN~2012cu de-reddened, we further demonstrate that in the case of SN~2012cu and SN~2011fe the twinning approach can provide highly precise and accurate relative distance measurements, as \citet{fakhouri15a} found.


This paper is organized as follows.  We present our spectroscopic observations of SN 2012cu in \S~\ref{sec:observations}.  In \S~\ref{sec:RV-ISM}, we measure \EBV, \RV, \AV, and the relative distance between SN~2012cu and SN~2011fe, using data from 10 epochs between \cuPzero and \cuPeleven days \edd{measured relative to the DayMax parameter of the SALT lightcurve fitter \citep{guy07a, guy10a}}.  We explore different approaches for handling spectral mismatches that remain.  We also measure the EWs of Na~I and a diffuse interstellar band (DIB) feature at 5780 \AA\ \edd{and examine whether these feature exhibit time variability.}  In \S~\ref{sec:discussion}, we determine the distance to the host galaxy of SN~2012cu and discuss the nature of foreground dust in its host.  We present our conclusions in \S~\ref{sec:conclude}. 

%
%

\newpage
\section{Observations} \label{sec:observations}

SN 2012cu was discovered on 2012, June 11.2 UT \citep{itagaki12a, ganeshallingam12a} and was classified as a Type~Ia on 2012, June 15 UT \citep{marion12a, zhang12a}.   NGC~4772 is the host, at a heliocentric redshift of $z = 0.003469 \pm 0.000017$ \citep{devaucouleurs91a}.
 
\begin{figure*}[!h]   
\centering
\includegraphics[width=.8\linewidth]{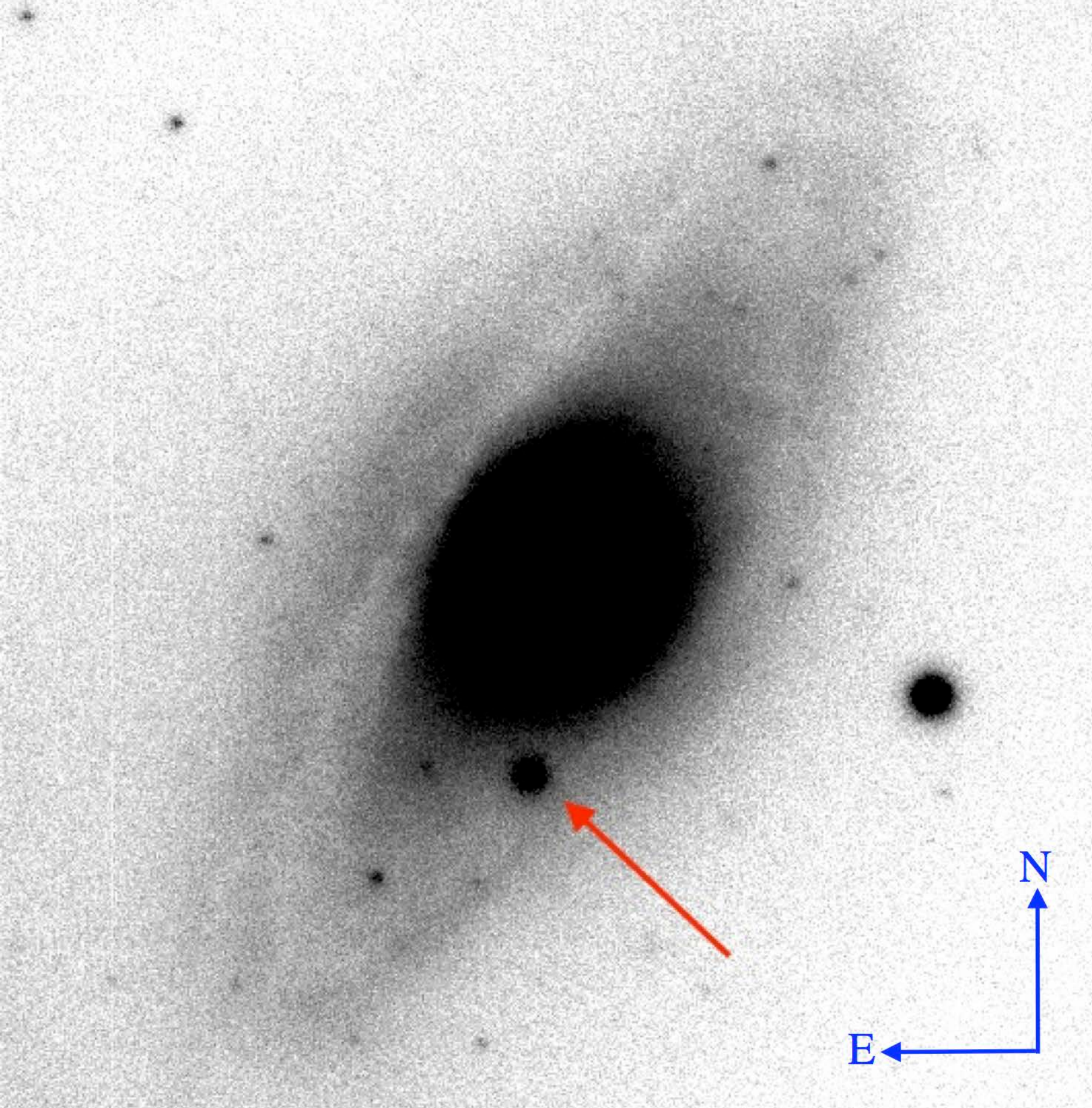}


\begin{picture}(0,0)
\put(-200, 325){\includegraphics[height=4cm]{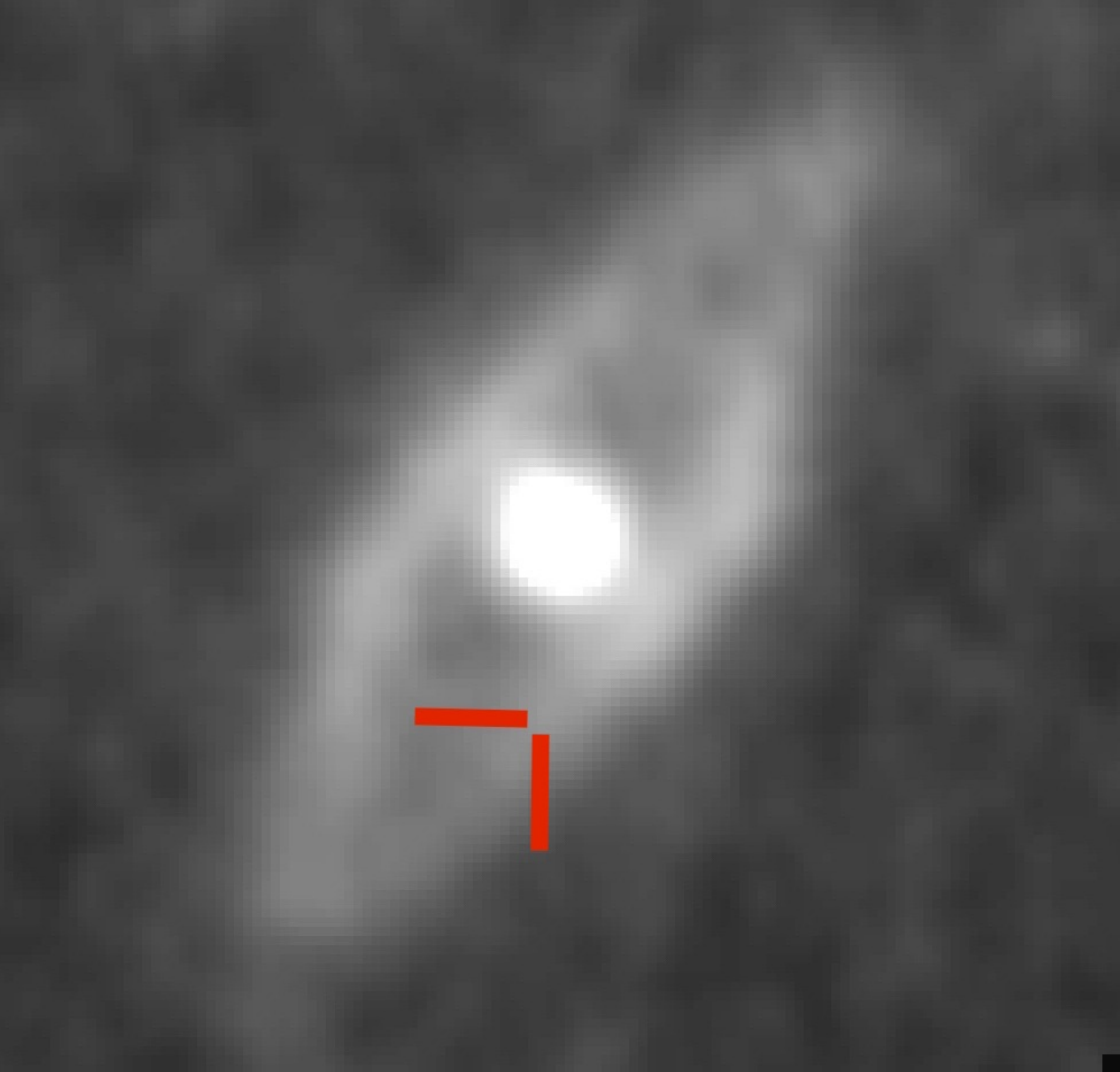}}
\end{picture}
\begin{picture}(0,0)
\put(91, 325){\includegraphics[height=4cm]{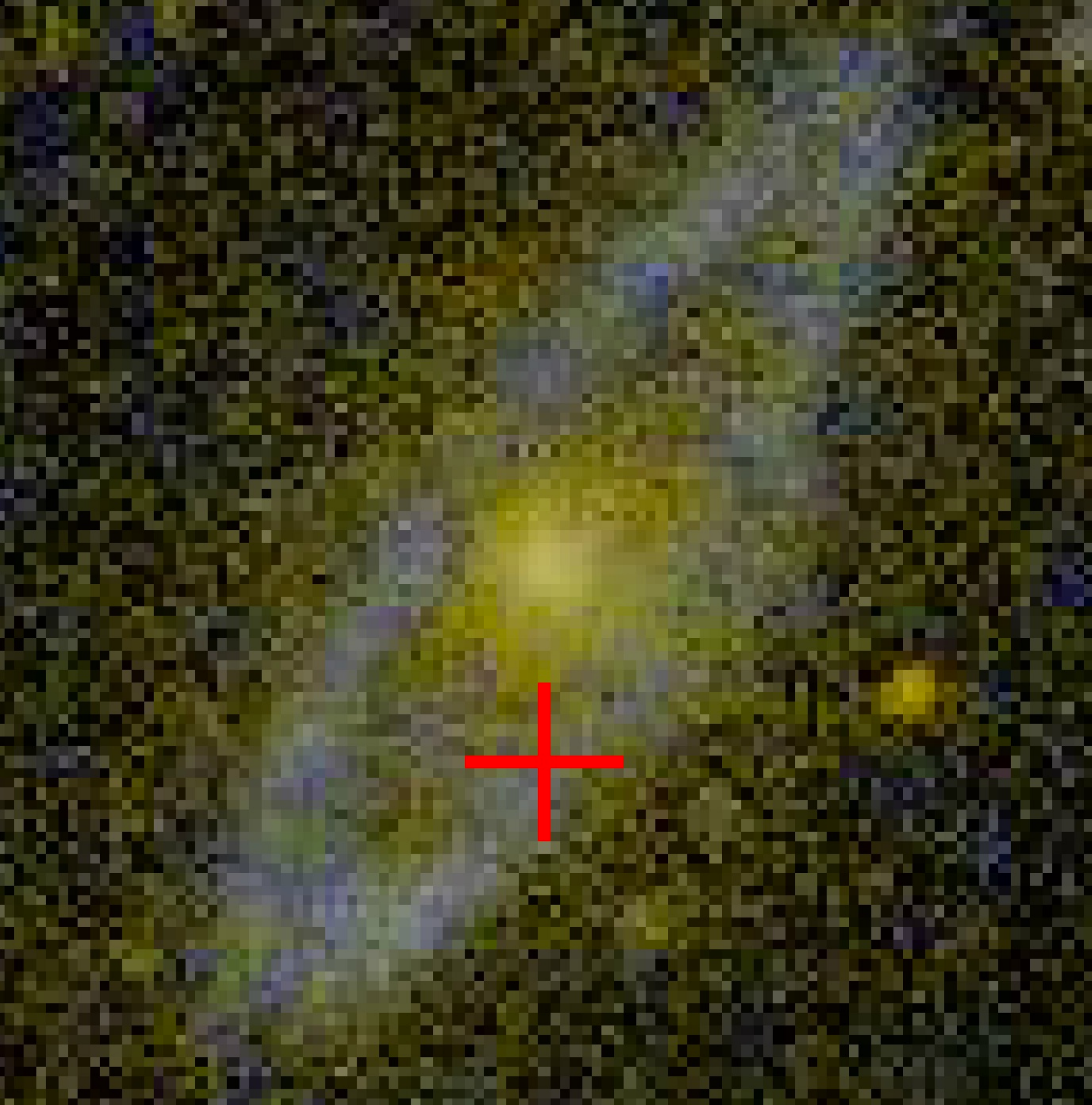}}
\end{picture}

\caption{\label{fig:acq-image-WISE}
$V$-band image of SN 2012cu (arrow) acquired using SNIFS.  \rf{The compass arrows are $20\twopr$ long.}  The inset at the upper left is the WISE 12-$\mu m$ image of the host galaxy NGC 4772, with the approximate projected position of SN~2012cu marked.  Both images clearly show the presence of a dust ring, in absorption in $V$-band and in emission at 12 $\mu$m.  \ed{The inset at the upper right is the GALEX 
image that shows the young stars along the dust lane in NGC~4772, with the crosshairs marking the position of SN~2012cu.}  
}
\end{figure*}

Figure~\ref{fig:acq-image-WISE} shows our $V$-band image of SN 2012cu and the host galaxy, NGC 4772.  The upper left inset is an image from the Wide-Field Infrared Survey Explorer \citep[WISE,][]{wright10a} in the 12-$\mu$m band,  
which \edd{spans the wavelengths} of some of the emission features of polycyclic aromatic hydrocarbons (PAHs) associated with dust \citep{li01a}.  The upper right inset is a GALEX \citep{bianchi14a} image that highlights the young stars along the dust lane.
The comparison of these images shows that SN 2012cu 
is projected onto a dust ring.  \ed{There is a reasonably high probability that \sncu is behind or embedded in the dust lane, and that it is mostly reddened by this dust detected in the interstellar medium (ISM) of the host galaxy.}

Spectra of SN 2012cu, spanning 17 epochs from \cuPzero to \cuPsixteen days.
were obtained by the SNfactory collaboration with its SuperNova Integral Field Spectrograph  \citep[SNIFS,][]{aldering02a, lantz04a} on the University of Hawaii 2.2-m telescope on Mauna Kea.  For the wavelength range of 3,300 - 9,200 \AA, the spectra of SN 2012cu are flux calibrated \citep{buton13a}, host-galaxy subtracted \citep{bongard11a}, corrected for Milky Way extinction \citep{schlafly11a}\ed{, and then de-redshifted to the rest frame for subsequent analysis.  Further details, including the observing log, a figure showing the entire the spectral time series, and information on how to access these data are provided in Appendix~\ref{sec:appendixB}.}



%
%

\footnotetext[1]{http://ned.ipac.caltech.edu/}

\newpage
\section{Dust Extinction Properties} \label{sec:RV-ISM}

\ed{In this section, we first describe how we fit for the dust extinction curve and the distance modulus difference (\Delmu) between SN~2012cu and SN~2011fe (\S~\ref{sec:Rv}).  In this effort, we explore different de-weighting schemes for regions where spectral differences are present.  We then turn our attention to Na I and diffuse interstellar band (DIB) features in the SN~2012cu spectra (\S~\ref{sec:ISM}).  Dust extinction curves are parametrized by quantities denoted as \EBV and \RV; these are not required to match the values that would be measured only using the $B$ and $V$~bands.} 

\subsection{Extinction Curve and Distance Modulus Difference}\label{sec:Rv}


\subsubsection{Simultaneously Fitting for \EBV, \RV and \Delmu}\label{sec:straight-binned-fit}

The best-fit SALT2.4 parameters for SN 2012cu are \edd{first determined to be} $(x_1^\prime, c^\prime)  = (0.021\pm0.123, 0.971\pm0.030)$ and \edd{$\mathrm{DayMax}^\prime = \mathrm{MJD} \, 56104.89 \pm 0.15$}.
\edd{Given the extreme color of SN~2012cu, much higher than the typical reddening in the training data for SALT, we apply an initial de-reddening using the extinction curve of \citet{fitzpatrick99a} 
and perform the fit again. We find $x_1  = -0.269\pm0.078$ and $\mathrm{DayMax} = \mathrm{MJD} \, 56105.06 \pm 0.11$.  The DayMax value has shifted by +0.16 days, and corresponds to June~27.1~UT.  To be consistent, we perform the SALT2.4 fit for SN~2011fe as well, and obtain $(x_1, c) = (-0.280 \pm 0.093, -0.058 \pm 0.027)$ and $\mathrm{DayMax} =  55815.20 \pm 0.08$.  (Our $x_1$, $c$, and DayMax are in good agreement with the corresponding values in \citet{pereira13a}, using SALT2.2.)  
The $x_1$ parameter is now very similar for these two SNe.}  
We further determine that this pair has a near-maximum twinness ranking = 0.1, on a scale from 0 to 1, placing the pairing of these two SNe among the best spectroscopic ``twins" \citep[][Figure 5]{fakhouri15a}.  \ed{It is even the case that both SNe exhibit the rare C II feature \citep{thomas11a}, \edd{albeit at higher velocity for SN~2012cu at the earliest phase,} strengthening the case for the similarity of the explosion physics.}

As SN 2011fe suffers nearly no host extinction, 
\edd{ for the rest of this paper we treat its spectra, which have been corrected for mild Milky Way (MW) extinction (\EBV = 0.0088 mag) by \citet{pereira13a}, as if they were un-reddened versions of SN 2012cu spectra.}

To determine the effects of dust reddening for SN 2012cu, we first pair the SN~2011fe spectra, also obtained with SNIFS \citep{pereira13a}, with those of SN~2012cu for cases where the phase difference is less than one day.  This is possible for 10 SN~2012cu phases between \cuPzero and \cuPeleven days, thanks to the dense temporal sampling of the \citet{pereira13a} data set (see Table~\ref{tab:phase-diff-meas-error-params}).  
\edd{\edd{In this paper, the phases for both SNe are always measured relative to their respective DayMax.}  The phase differences (0.2 -- 0.3 days; see Table~\ref{tab:phase-diff-meas-error-params}) are somewhat larger than the combined phase uncertainty between the two SNe (0.14 days).  As a test, we have performed interpolation for several phases, including for phase pairings with a 0.3 d phase difference, and find these phase differences make very little difference for the best-fit results.  Thus there is no need to interpolate.}

For these 10 pairs of spectra, we divide the fluxes evenly into 925 bins above 3600~\AA; between~3300 \AA\ and 3600~\AA\ we use \edd{two} bins due to the lower S/N here for the highly \ed{reddened} SN 2012cu.  Therefore there are a total of \edd{927} bins.  We use these same bins for both SNe.

\ed{The extinction curve models of \citet{cardelli89a}, \citet{odonnell94a}, and \citet{fitzpatrick99a} (CCM89, OD94, and F99, respectively) are the most commonly used in the literature.  All three are parameterized by \RV and \EBV.  F99 pointed out the tension between observation and CCM89 in the wavelength range that roughly corresponds to the $i$ and $r$ bands.  This has been subsequently confirmed by other authors, not just for CCM89 but also for OD94 \citep[e.g.][]{schlafly10a, mortsell13a}.  A more detailed discussion is given in Appendix A.  For the remainder of this section, we will use F99.}

As we will show below, despite being good ``twins", \ed{the high S/N ratio of the SNIFS data makes the differences in spectral features between SN 2011fe and SN 2012cu clearly visible.  Such differences are stronger for pre-maximum phases}.  \ed{We compare three approaches toward the regions where the spectral differences are high.  In Approach~I, we will treat all wavelength bins in the same manner.  In Approach~II, we will identify and de-weight these regions appropriately.  Finally in Approach~III, we will discard these regions altogether in the fitting process.}

\newpage
\begin{deluxetable}{cccccccc}
\tablecaption{\label{tab:phase-diff-meas-error-params} Phase Comparison and Best-Fit Parameters Using Approach~I}
\tablehead{\colhead{Pair ID} & \colhead{SN 2012cu} & \colhead{SN 2011fe} & \colhead{$E(B-V)$} & \colhead{$R_V$} & \colhead{$\Delta \mu$} & \colhead{$A_V$} & \colhead{$\chi^2_{\nu}$}\\ \colhead{ } & \colhead{Phase (days)} & \colhead{Phase (days)} & \colhead{ } & \colhead{ } & \colhead{ } & \colhead{(= $R_V \cdot E(B-V)$)} & \colhead{ }}
\startdata
1 & $-6.8$ & $-7.0$ & $0.995\pm0.005$ & $2.908\pm0.024$ & $1.964\pm0.013$ & $2.895\pm0.015$ & $51.22$ \\
2 & $-1.7$ & $-2.0$ & $0.989\pm0.006$ & $3.082\pm0.025$ & $1.873\pm0.012$ & $3.048\pm0.014$ & $88.48$ \\
3 & $3.3$ & $3.0$ & $1.019\pm0.006$ & $2.987\pm0.024$ & $1.868\pm0.012$ & $3.043\pm0.013$ & $99.05$ \\
4 & $6.3$ & $6.0$ & $1.017\pm0.006$ & $2.962\pm0.023$ & $1.864\pm0.011$ & $3.012\pm0.013$ & $53.41$ \\
5 & $8.3$ & $8.0$ & $1.010\pm0.007$ & $2.970\pm0.024$ & $1.884\pm0.011$ & $3.001\pm0.012$ & $22.70$ \\
6 & $11.2$ & $11.0$ & $0.982\pm0.006$ & $2.992\pm0.024$ & $2.043\pm0.010$ & $2.939\pm0.012$ & $20.49$ \\
7 & $16.2$ & $16.0$ & $0.911\pm0.008$ & $3.218\pm0.033$ & $2.016\pm0.011$ & $2.931\pm0.013$ & $27.47$ \\
8 & $18.2$ & $18.0$ & $0.907\pm0.009$ & $3.270\pm0.038$ & $1.968\pm0.013$ & $2.966\pm0.015$ & $57.26$ \\
9 & $21.2$ & $21.0$ & $0.912\pm0.011$ & $3.265\pm0.041$ & $2.027\pm0.012$ & $2.976\pm0.014$ & $53.03$ \\
10 & $23.2$ & $23.0$ & $0.915\pm0.011$ & $3.177\pm0.041$ & $1.998\pm0.012$ & $2.908\pm0.014$ & $26.77$ \\
\enddata
\end{deluxetable}

\edd{For Approach~I, when we simultaneously fit for 
$E(B-V)$, $R_V$, and $\Delta\mu$, only measurement uncertainties are included.}
For each phase the fit must de-redden the spectrum of SN 2012cu using the dust extinction curve of F99, calculate the AB magnitude \citep[see, e.g.,][]{bessell12a} for each bin, $m_{AB, \, k}^\mathrm{12cu, \, dereddened}(E(B-V), R_V)$, which can be more conveniently written as $m_{AB, \, k}^\mathrm{12cu} + 2.5\,\mathrm{log}(F99(E(B-V), R_V))$, and then compare the results with \snfe.  For each separate matching phase, we therefore minimize

\begin{equation}
\chi^2 =  \sum\limits_{k} {\frac {\left[ m_{AB, \, k}^\mathrm{12cu} - m_{AB, \, k}^\mathrm{11fe} + 2.5\,\mathrm{log}(F99(E(B-V), R_V)) - \Delta \mu\right]^2} {\sigma^2_{k, \mathrm{meas}}}}
\label{eqn:chi2_cleaner},
\end{equation}

\noindent
where $\sigma_{k, \mathrm{meas}}^2 = \left(\sigma^\mathrm{11fe}_{AB, \, k}\right)^2 +  \left(\sigma^\mathrm{12cu}_{AB, \, k}\right)^2$ is the total measurement variance for bin $k$ in magnitude space.  \ed{SN 2012cu has a wavelength-independent (``gray") scatter of 0.025 mag, uncorrelated between phases \citep{buton13a}.  For SN~2011fe it is between 0.03 -- 0.06 mag, again uncorrelated between phases.  These larger values stem from the unusually high airmasses at which some \snfe observations were taken.  Effectively this gray scatter is absorbed into the uncertainty for \Delmu and does not affect $\chi^2$.}

In Figure~\ref{fig:2panel-spect-comp} we show the binned spectrum of SN 2012cu at phase \cuPzero days in units of AB magnitude, dereddened with the best-fit \EBV and \RV and vertically shifted by the best-fit \Delmu, together with the spectrum of SN 2011fe for the corresponding phase.  The minimum reduced $\chi^2$ is $\chi^2_\nu = 51.22$.  \ed{It is evident that this high value of \chisqnu is due to spectral feature differences between the two SNe.  In the next section, we will present our main approach to address the unaccounted-for uncertainty.  Here, to set a baseline for comparison with Approaches~II and III, we simply uniformly inflate the errors so that \chisqnu is rescaled to 1.  This does not change best-fit values or the shape of confidence contours.  The error bars on the fitting parameters in the top panel are those computed after this rescaling.}
The vertical green bands are the regions of the \edd{highly variable} Si~II and Ca~II features \edd{identified} in \citet{chotard11a}, which will be henceforth referred to as the ``Chotard regions."  Outside of these regions, some of the other well-known SN spectral features are also labeled.  The sharp feature around 5896~\AA\ labeled in red is Na I D absorption.  The bottom panel shows the residual of the fit weighted by the measurement uncertainties, or the ``pull spectrum."  It is clear that the two spectra can be very different outside of the Chotard regions, as well.  As examples (see Figure~\ref{fig:2panel-spect-comp}), they are significantly different for: (i) the regions of Fe II $\lambda\lambda$5018, 5169, S II $\lambda$5640, and O I $\lambda$7773, (ii) the emission part of the P-Cygni profile of one of the Chotard Si II features, centered around 6350~\AA, and (iii) the emission part of the Ca II IR triplet, centered around 8600~\AA.  

\noindent%
\begin{minipage}{\linewidth}
\makebox[\linewidth]{
  \includegraphics[keepaspectratio=true,scale=0.4]{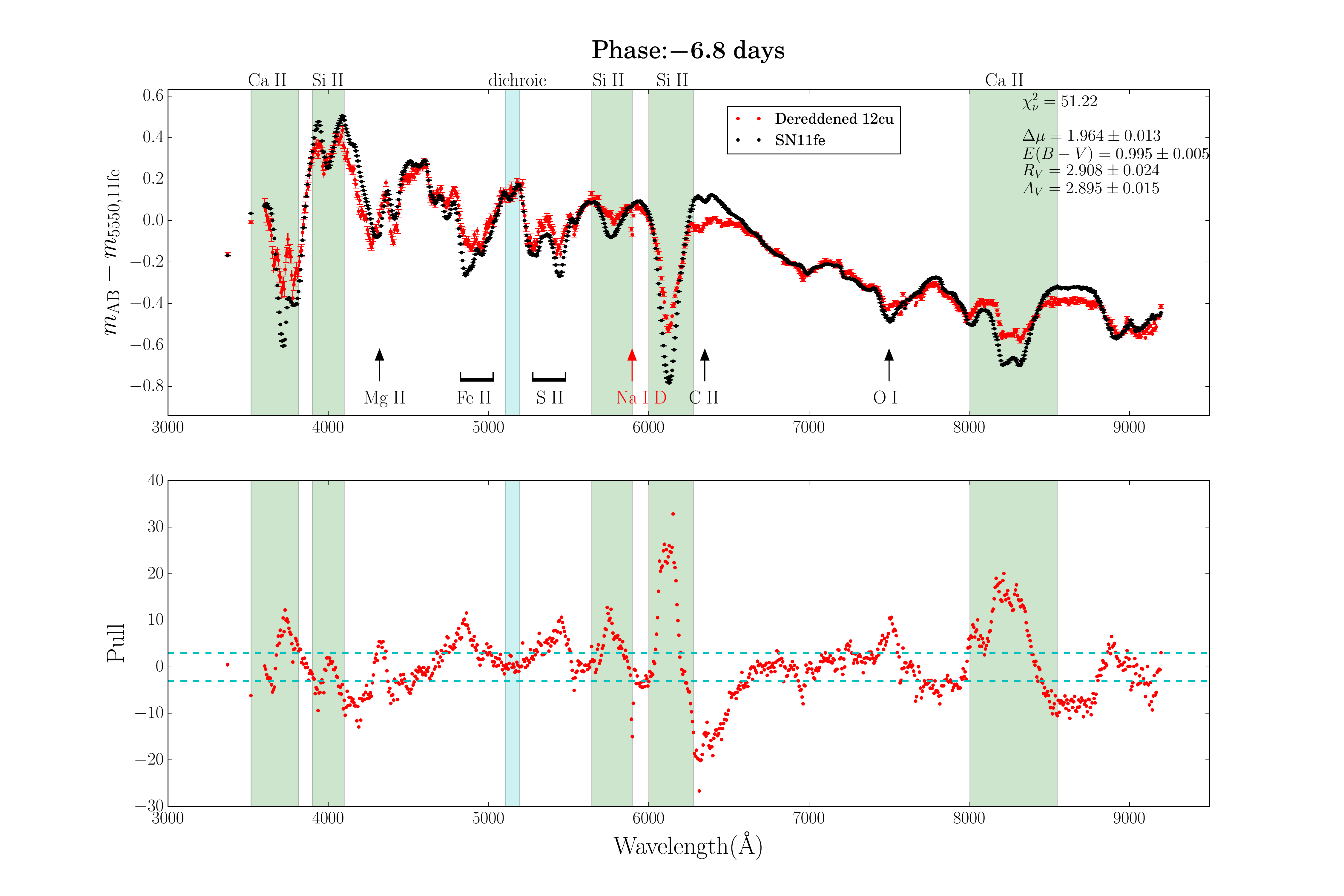}}
\captionof{figure}{\textbf{Top Panel} shows the binned SN 2012cu spectrum (red) in units of AB magnitude for the phase of \cuPzero days dereddened with the best-fit \EBV and \RV, and shifted vertically by the best-fit \Delmu, for comparison with the template spectrum of SN~2011fe (black) \ed{for the same phase}.  \rf{The zero point is arbitrary, chosen to be the magnitude at 5550 \AA\ for \snfe.}  Even for well-twinned SNe, it is for pre-maximum phases that spectral differences are more likely, so this pair is among those with the largest level of disagreement.  Very small differences can be seen here given the high S/N ratio of these spectra.  For comparison at later phases, see Figure~\ref{fig:all-phases-orig}.  The vertical green bands are the regions of the Si~II and Ca~II features found to be highly variable between SNe in \citet{chotard11a}.  Outside of these regions, some of the other well-known SN spectral features are also labeled, as well as the Na~I~D absorption feature.  \textbf{Bottom Panel} shows the difference in magnitude between the two spectra in the top panel (the residual), weighted by measurement uncertainty, or the ``pull."  The peaks and valleys correspond to where the spectra are the most different.  The cyan dashed lines indicate 3$\,\sigma$ deviation.  It is clear that the high value of \chisqnu is due to spectral feature differences between the two SNe, despite the fact that they are good matches relative to other SNe.  While the Chotard regions capture the greatest spectral variations, the two spectra can nonetheless be very different outside of these regions.  
}\label{fig:2panel-spect-comp}
\end{minipage}

\ed{In Figure~\ref{fig:EBV-RV-contour_phase0}, we present the confidence contours for \EBV and \RV, after rescaling \chisqnu to 1, which is equivalent to inflating the variances in Equation~\ref{eqn:chi2_cleaner} uniformly across all wavelengths}, and marginalizing over the distance modulus difference \Delmu.  \EBV and \RV are clearly anti-correlated; this is true for all 10 phases.  The uncertainties for the extinction, \AV (simply the product of the best-fit \EBV and \RV), are computed by taking into account the covariance between these two quantities.  It is notable that, owing to this anti-correlation, the uncertainty for \AV is smaller than for \RV.

\noindent
\begin{minipage}{\linewidth}
\makebox[\linewidth]{
  \includegraphics[keepaspectratio=true,scale=0.6]{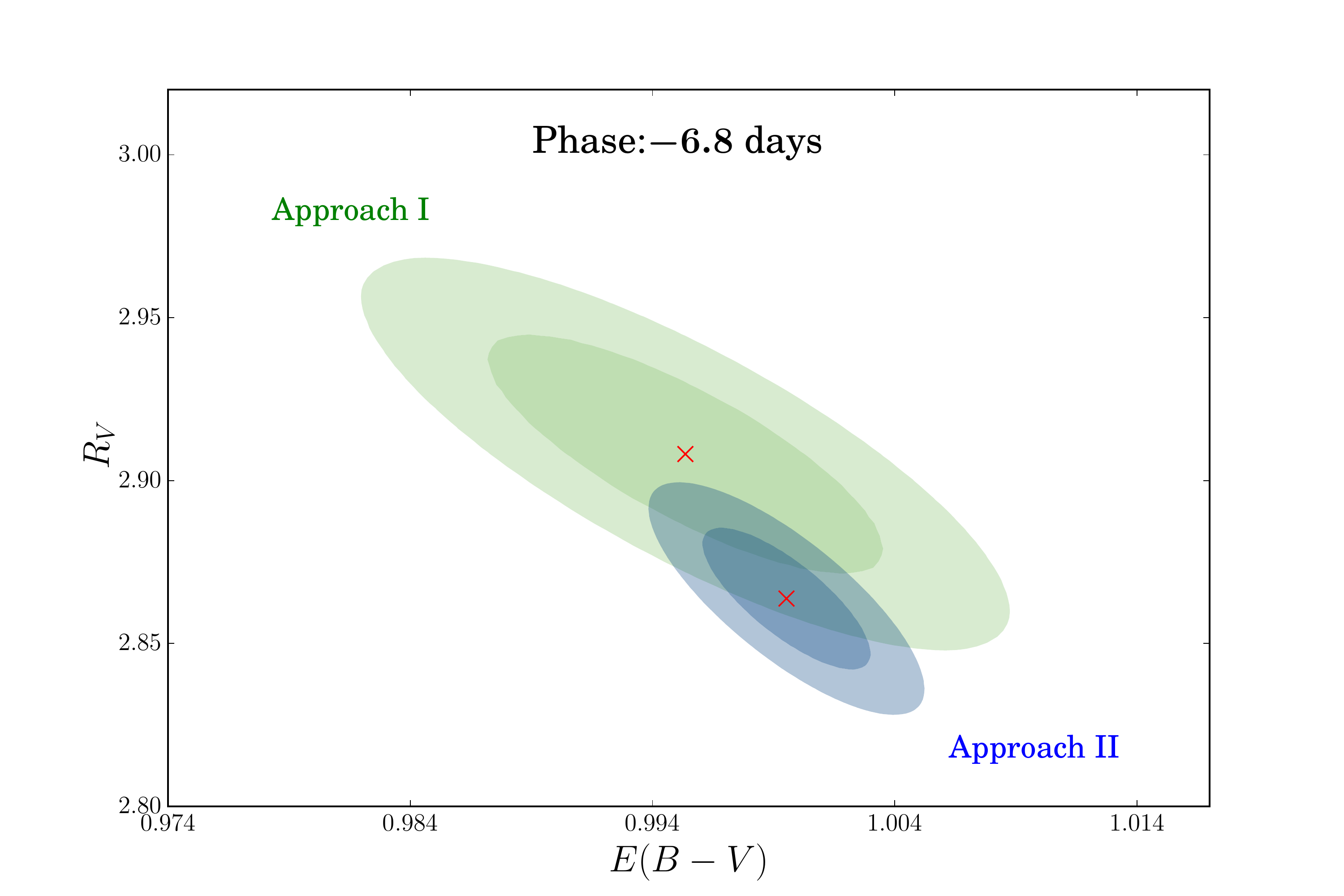}
} \captionof{figure}{\ed{The green contours indicate the 1 and 2$\,\sigma$ confidence regions for \EBV and \RV for the phase of \cuPzero days for Approach~I (see text).  The cross marks the best-fit values.  Likewise, the blue contours are for Approach~II.  All 10 phases between \cuPzero and \cuPeleven days exhibit a similar anti-correlation between these two fitting parameters.}
}\label{fig:EBV-RV-contour_phase0}  
\end{minipage}

\noindent
\begin{minipage}{\linewidth}
\makebox[\linewidth]{
  \includegraphics[keepaspectratio=true,scale=0.25]{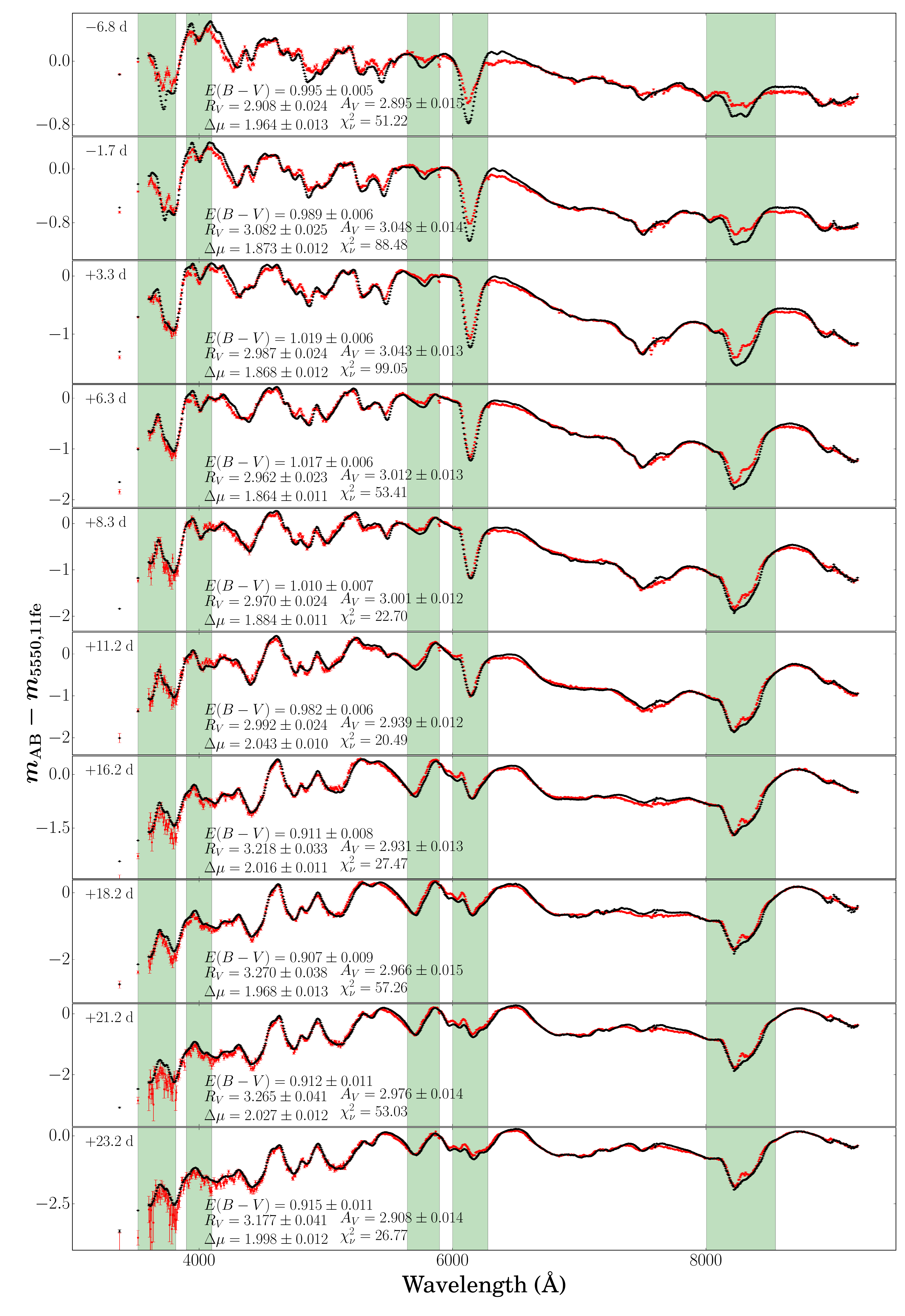}}
\captionof{figure}{Full SN~2012cu spectrophotometric time series used in the analysis.  The phases are shown in the upper left corner of each panel.  Binned SN~2012cu spectra (red) in units of AB magnitude dereddened with the best-fit \EBV and \RV, and shifted vertically by the best-fit \Delmu, are shown together with spectra of SN~2011fe with corresponding phases.  As with Figure~\ref{fig:2panel-spect-comp}, the vertical green bands are the Chotard regions.
}\label{fig:all-phases-orig}  
\end{minipage}

We apply the same fitting procedure to each of the 10 phases.  The results are shown in Figure ~\ref{fig:all-phases-orig}.  Larger values of \chisqnu are, generally, associated with the earlier phases, owing to their greater spectral diversity.  We will address this issue further in the next subsection, \S\ref{sec:spectral-features-all}.  The phase-by-phase summary for the best-fit \EBV, \RV, \Delmu, and the derived quantity, \AV, are presented in Table~\ref{tab:phase-diff-meas-error-params}.  
  The uncertainty-weighted average and RMS values across the ten phases (Table~\ref{tab:phase-diff-meas-error-params}), are (\EBV, RMS) = (\numprint{\EBVI}, \numprint{\eEBVI})~mag, (\RV, RMS) = (\numprint{\RVI},  \numprint{\eRVI}), (\Delmu, RMS) = (\numprint{\DelmuI},  \numprint{\eDelmuI})~mag, and (\AV, RMS) = (\numprint{\AVI},  \numprint{\eAVI})~mag.  \edd{Conservatively, for \RV and \EBV, we take the uncertainties on the means to be the RMS dispersions.  The uncertainty on \Delmu is discussed in \S~\ref{sec:host-dist}, as it requires further consideration.}


\subsubsection{Effects of Spectral Feature Differences}\label{sec:spectral-features-all}

In the previous section we showed that even though SN 2012cu and SN 2011fe are good ``twins,'' there are significant spectral differences between them.  In this section, by presenting the results of \edd{two alternative ways to handle these spectral feature differences}, we will show that they do \textit{not} have a significant effect on the best-fit extinction and relative distance parameters.

\ed{In Approach~II, we de-weight SN spectral feature differences but leave differences due to measurement noise intact.}  To identify where the spectra behave differently for SN 2011fe and SN 2012cu the terms in the summation in Equation~\ref{eqn:chi2_cleaner} are plotted against the central wavelengths of the corresponding bins.  As shown in Figure~\ref{fig:2panel-spect-comp}, in the ($residual/\sigma$) spectrum, or pull spectrum, the locations of the peaks and valleys are where the spectra of 2011fe and 2012cu are most different.  
We convolve the pull spectrum
with a Gaussian kernel and through trial and error find that a Gaussian with a standard deviation of 25 \AA\ captures the spectrally correlated differences (features) discernible by eye (Figure~\ref{fig:3panel-spect-comp_convl}).  We now use the result of the convolution, $p_k$, to de-weight the spectral differences.  \ed{First we set $p_k$ to zero for the wavelength bins where $|p_k| \leq 3$.}  We de-weight the complementary bins, where $|p_k| > 3$, by rescaling $p_k$ in such a way that when added in quadrature to the denominator in Equation~\ref{eqn:chi2_cleaner} the reduced $\chi^2$ for these regions is 1.  \edd{The spectral regions that are not de-weighted span the full wavelength range.}  We then add the rescaled $p_k^2$ to $\sigma^2_{k, \mathrm{meas}}$ in the denominator of Equation~\ref{eqn:chi2_cleaner}.  In order to achieve \edd{an average} \chisqnu~$\sim 1$ across all phases, \ed{we find that for the wavelength regions where $|p_k| \leq 3$, it is necessary to add an error floor of $\sigma_\mathrm{floor} = \ed{0.03}$ mag.  This error floor is \edd{treated as} an uncorrelated error, and is not related to the gray scatter mentioned in \S\ref{sec:straight-binned-fit}.}  Thus Equation~\ref{eqn:chi2_cleaner} becomes

\begin{minipage}{\linewidth}
\begin{equation}
\chi^2 =  \sum\limits_{k} {\frac {\left[ m_{AB, \, k}^\mathrm{12cu} - m_{AB, \, k}^\mathrm{11fe} + 2.5\mathrm{log}(F99(E(B-V), R_V)) + \Delta \mu\right]^2} {\sigma^2_{k, \mathrm{meas}} + p_k^2 + \sigma^2_\mathrm{floor}}}
\label{eqn:chi2-full-var}.
\end{equation}
\end{minipage}

Using Equation~\ref{eqn:chi2-full-var}, we once more perform $\chi^2$ minimization.  In Figure~\ref{fig:3panel-spect-comp_convl} we show the best-fit de-reddened SN 2012cu spectrum at \cuPzero days, vertically shifted by the best-fit \Delmu, together with the 2011fe spectrum \ed{at the corresponding phase}.  We obtain \chisqnu = \ed{0.90}, close to 1 as expected.  \ed{We note the pull values (third panel in Figure~\ref{fig:3panel-spect-comp_convl}) for $\lambda \lesssim 6000$ \AA\ are slightly larger than for the longer wavelengths because the measurement uncertainties for the bluer wavelengths are larger due to extinction and therefore a uniform error floor doesn't have as much of an effect.  However, we will show below that the particular choice of weighting scheme doesn't affect the best-fit results in a significant way.}  We then apply the same fitting procedure to each of the 10 phases. The results are presented in the left panels of Figure~\ref{fig:all-phases-convl} and Table~\ref{tab:spect-diff-params}.  The values of \chisqnu now range from \edd{0.78 to 1.24.}  The right panels of Figure~\ref{fig:all-phases-convl} show the convolution result for the pull spectrum for each of the 10 epochs.

\noindent
\begin{minipage}{\linewidth}
\makebox[\linewidth]{
  \includegraphics[keepaspectratio=true,scale=0.4]{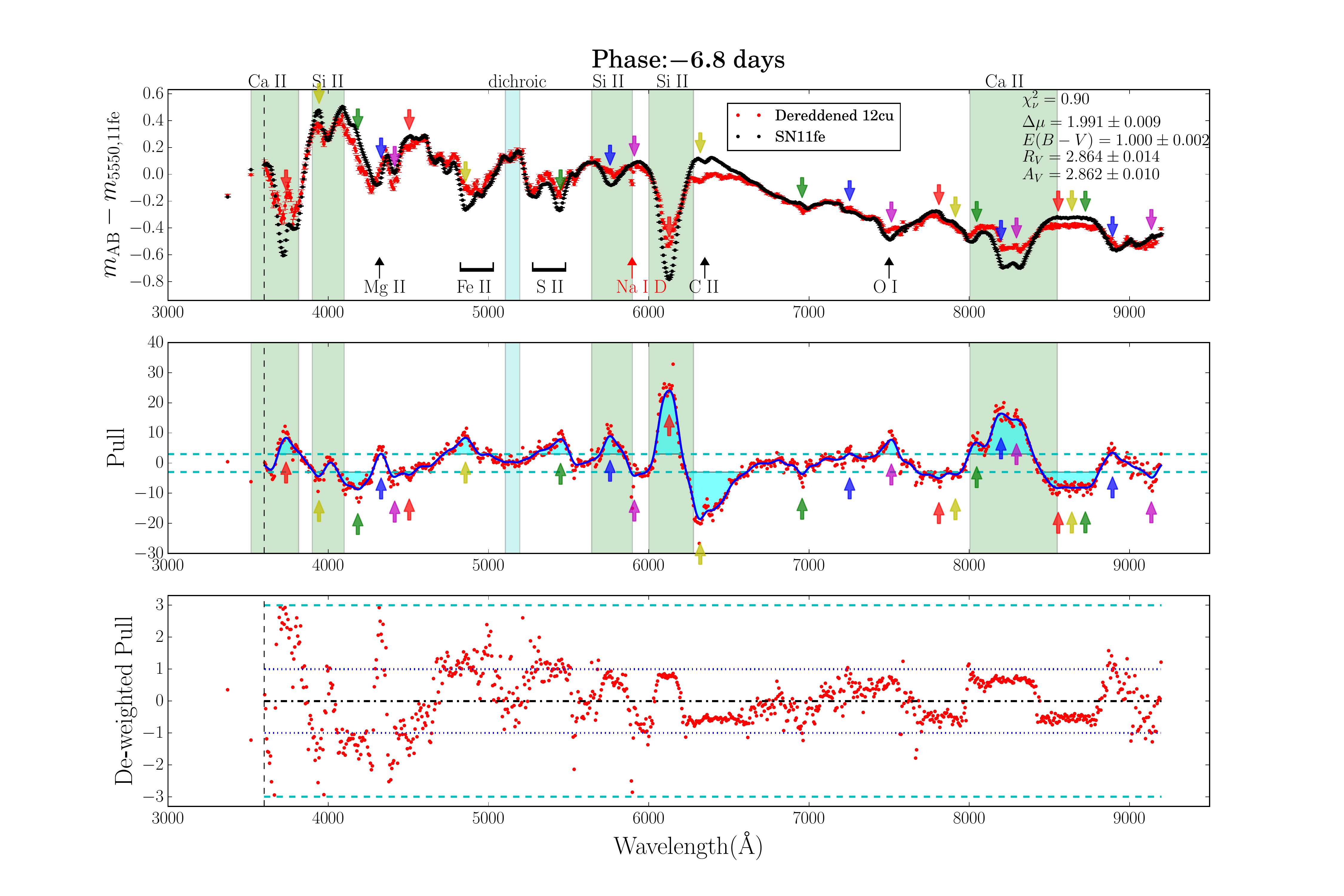}}
\captionof{figure}{This figure demonstrates the technique used in Approach II, to de-weight the regions where the spectral differences between SN 2011fe and SN2012cu are large.  \textbf{Top Panel}: The best-fit result after our technique has been applied, achieving \chisqnu~$\sim1$ as expected.  
\textbf{Middle Panel}: This panel demonstrates the de-weighting used in this technique.  The pull spectrum shown here is the same as the one in the middle panel of Figure~\ref{fig:2panel-spect-comp}, \rf{i.e., before de-weighting}.  The solid blue line is the result of the Gaussian convolution (see text).  Each arrow points to an extremum whose absolute value is greater than 3\,$\sigma$, marked by the cyan dashed lines.  The arrows are matched by color between the top and middle panels and demonstrate that spectral feature differences are identified by this technique.  The filled cyan regions are the wavelength regions where the deviation is greater than 3$\,\sigma$.  To de-weight these regions, an appropriately scaled quantity, $p_k$ (see text), is added in quadrature to the measurement uncertainty in Equation~\ref{eqn:chi2_cleaner} so that the \chisqnu for these cyan regions is 1.  In addition we add a \ed{0.03 mag error floor to the complementary wavelength regions, where the deviation $\leq 3\,\sigma$,} to achieve \edd{an average} \chisqnu$\sim1$ across all phases (see Top Panel).  \textbf{Bottom Panel}: The pull spectrum after the application of our technique (\rf{i.e., after de-weighting}).  The blue dotted and cyan dashed lines indicate 1 and 3\,$\sigma$ deviations, respectively.
}\label{fig:3panel-spect-comp_convl}
\end{minipage}

The weighted average and RMS values across the 10 phases are (\EBV, RMS) = ($\numprint{\EBVII}, \numprint{\eEBVII}$), (\RV, RMS) = ($\numprint{\RVII}, \numprint{\eRVII}$), (\Delmu, RMS) = ($\numprint{\DelmuII}, \numprint{\eDelmuII}$), and (\AV, RMS) = ($\numprint{\AVII}, \numprint{\eAVII}$).  These results, along with those from Approach~I, are summarized in Table~\ref{tab:summary-3-approaches} and Figure~\ref{fig:params-phases-summary}.  Once again we take the uncertainties on the means for \edd{\RV and \EBV} to be the RMS dispersions across phases.    
 
\newpage
\noindent
\begin{minipage}{\linewidth}
\makebox[\linewidth]{
  \includegraphics[keepaspectratio=false,scale=0.21]{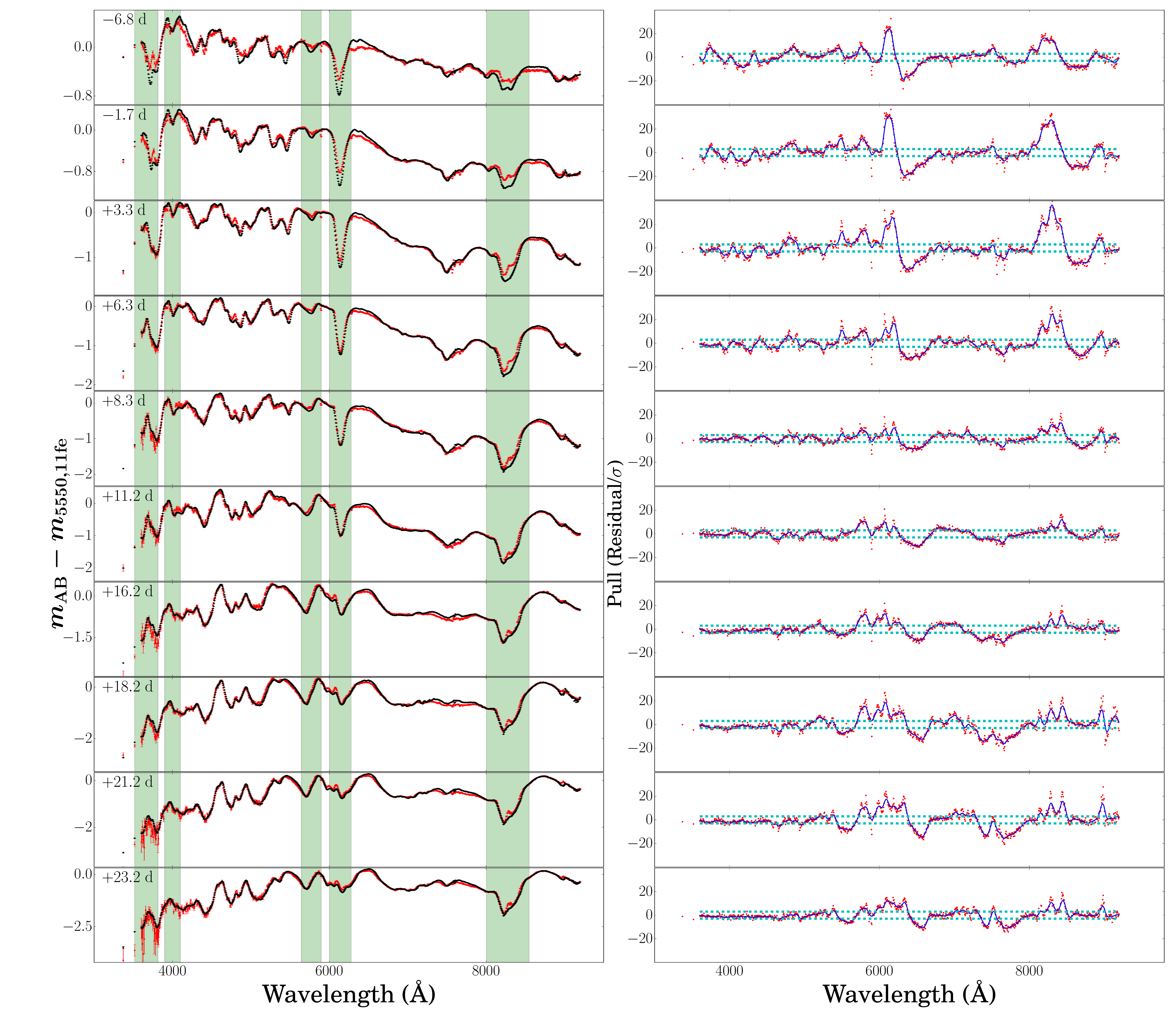}}
\captionof{figure}{The left panels are the same as Figure~\ref{fig:all-phases-orig}, except the fitting process now is that of Approach II, which de-weights the wavelength regions where there are significant spectral feature differences (see text).  The results of convolving the pull spectrum for each phase with a Gaussian kernel are shown in the right panels.  The cyan dashed lines indicate 3$\,\sigma$ deviations.  
}\label{fig:all-phases-convl}  
\end{minipage}

We have pursued one more approach.  In Approach~III, we increase $p_k$ to an arbitrarily high value, essentially removing data in the wavelength regions with pull values greater than 3 (the cyan regions in Figure~\ref{fig:3panel-spect-comp_convl})  from the fitting process.  The average values of our best-fit parameters across the 10 epochs are  (\EBV, RMS) = (\numprint{\EBVIII}, \numprint{\eEBVIII}), (\RV, RMS) = (\numprint{\RVIII},  \numprint{\eRVIII}), (\Delmu, RMS) = (\numprint{\DelmuIII}, \numprint{\eDelmuIII}), and (\AV, RMS) = (\numprint{\AVIII}, \numprint{\eAVIII}).   These results are \edd{very close} to those obtained in Approach~II.  The agreement among the results from these three approaches  ---  to well within their RMS fluctuations (Table~\ref{tab:summary-3-approaches})  ---  shows that our best-fit results are not sensitive to the particular scheme of de-weighting the regions of high spectral variation, including simply uniformly increasing the measurement uncertainties with no relative de-weighting (Approach~I).  \edd{However, the RMS values improve substantially when some reasonable form of de-weighting is applied to spectral features.}

\newpage
\begin{deluxetable}{cccccc}
\tablecaption{\label{tab:spect-diff-params} Spectral Features De-weighted Best-Fit Parameters Using Approach~II}
\tablehead{\colhead{SN 2012cu} & \colhead{$E(B-V)$} & \colhead{$R_V$} & \colhead{$\Delta \mu$} & \colhead{$A_V$} & \colhead{$\chi^2_{\nu}$}\\ \colhead{Phase (days)} & \colhead{ } & \colhead{ } & \colhead{ } & \colhead{(= $R_V \cdot E(B-V)$)} & \colhead{ }}
\startdata
$-6.8$ & $1.000\pm0.002$ & $2.864\pm0.014$ & $1.991\pm0.009$ & $2.862\pm0.010$ & $0.90$ \\
$-1.7$ & $0.998\pm0.002$ & $3.012\pm0.014$ & $1.912\pm0.009$ & $3.006\pm0.010$ & $0.83$ \\
$3.3$ & $1.038\pm0.002$ & $2.886\pm0.013$ & $1.918\pm0.008$ & $2.995\pm0.009$ & $0.77$ \\
$6.3$ & $1.033\pm0.002$ & $2.871\pm0.014$ & $1.909\pm0.009$ & $2.967\pm0.010$ & $0.88$ \\
$8.3$ & $1.019\pm0.003$ & $2.917\pm0.017$ & $1.913\pm0.010$ & $2.972\pm0.011$ & $1.22$ \\
$11.2$ & $0.987\pm0.003$ & $2.953\pm0.018$ & $2.063\pm0.009$ & $2.915\pm0.011$ & $1.24$ \\
$16.2$ & $0.947\pm0.003$ & $3.081\pm0.020$ & $2.038\pm0.009$ & $2.917\pm0.011$ & $1.20$ \\
$18.2$ & $0.953\pm0.003$ & $3.078\pm0.020$ & $2.008\pm0.010$ & $2.933\pm0.011$ & $1.02$ \\
$21.2$ & $0.966\pm0.004$ & $3.063\pm0.019$ & $2.058\pm0.009$ & $2.958\pm0.010$ & $0.99$ \\
$23.2$ & $0.967\pm0.004$ & $2.992\pm0.020$ & $2.027\pm0.009$ & $2.893\pm0.011$ & $1.17$ \\
\enddata
\end{deluxetable}

The average values for the fitting parameters determined by Approach~II in this subsection (\S\ref{sec:spectral-features-all}) are adopted as the best-fit host galaxy reddening (\EBV) and total-to-selective extinction ratio (\RV), and distance modulus difference between SN 2012cu and SN 2011fe (\Delmu).
\edd{This approach gives the lowest RMS on \AV and \Delmu.}  Our best-fit \EBV(= \nprounddigits{2}$\numprint{\EBVII}\pm\numprint{\eEBVII}$) compares well with the value \EBV = 0.99$\pm$0.03 reported by \citet{amanullah15a} based on broadband photometry from UV to NIR for SN~2012cu using the extinction curve of F99.  Our best-fit \RV(= $\numprint{\RVII}\pm\numprint{\eRVII}$) is slightly higher than their \RV = 2.8$\pm$0.1.  It is striking that optical spectrophotometry alone performs as well as combined UV, optical, NIR broadband photometry for this case. 


\noindent
\begin{minipage}{\linewidth}
\makebox[\linewidth]{
  \includegraphics[keepaspectratio=true,scale=0.65]{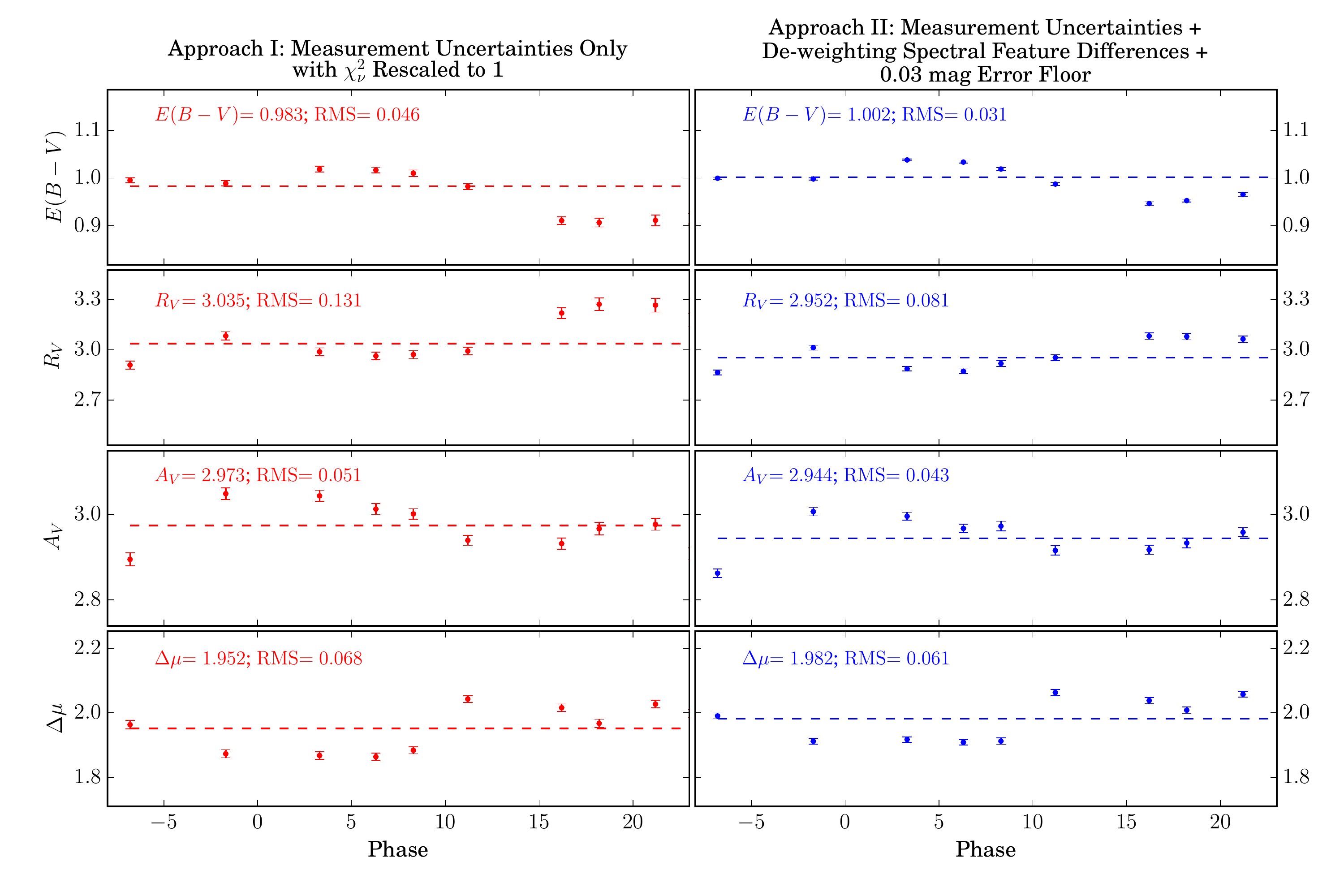}}
\captionof{figure}{\rf{The best-fit \RV, \EBV, \Delmu, and the derived quantity, \AV,} for the 10 phases of SN~2012cu between \cuPzero and \cuPeleven days.  
The weighted averages of these quantities are drawn as dashed lines and are printed \rf{in their respective panels}, along with the RMS dispersions.  \rf{\textbf{Left Column}:} Summary of best-fit results from Approach~I, where we only take into account measurement uncertainties, but rescale to \chisqnu~$\sim 1$ (see text) to determine the uncertainties of the fitting parameters for each phase.  \rf{\textbf{Right Column}:} Summary of best-fit results from Approach~II, where in addition to taking into account the measurement uncertainties, we de-weight the wavelength regions where the spectral differences are high (see text) and add a 0.03 mag error floor.  The results from these two approaches differ slightly and only past day 15.  Overall, Approach~II \edd{more than} halves the variance for \RV.  The anti-correlation across the phases between \RV and \EBV can be seen in both cases.  \rf{The RMS scatters seen here (and in Tables~\ref{tab:phase-diff-meas-error-params} and~\ref{tab:spect-diff-params}) are larger than what the per-phase error bars indicate.  We address this discrepancy in \S~\ref{sec:dust-nature}.}
}\label{fig:params-phases-summary}  
\end{minipage}

For Approach~II, the results for the last four epochs are half as far from the mean compared with Approach~I.  Since late epochs are where light echoes \citep[e.g.][]{patat05a} can become most prominent, this is an important improvement.



\nprounddigits{3}
\begin{deluxetable}{lcccc}
\tablewidth{0pt}
\tabletypesize{\scriptsize}
\tablecaption{\label{tab:summary-3-approaches} Comparison of Approaches~I, II, and III}
\tablehead{\colhead{Spectral Mismatch Approach} & \colhead{ \EBV} & \colhead{\RV} & \colhead{\AV}  & \colhead{\Delmu} \\
\colhead{} & \colhead{{mean} {RMS}} & \colhead{{mean} {RMS}} & \colhead{{mean} {RMS}} & \colhead{{mean} {RMS}}
}
\startdata
I. Not De-weighted (\S\ref{sec:straight-binned-fit})       & \numprint{\EBVI}  \numprint{\eEBVI}  & \numprint{\RVI}  \numprint{\eRVI}  & \numprint{\AVI}  \numprint{\eAVI} & \numprint{\DelmuI}  \numprint{\eDelmuI}  \\
II. De-weighted     (\S\ref{sec:spectral-features-all})    & \numprint{\EBVII}  \numprint{\eEBVII}  & \numprint{\RVII}  \numprint{\eRVII}  & \numprint{\AVII}  \numprint{\eAVII} & \numprint{\DelmuII}  \numprint{\eDelmuII} \\
III. Removed           (\S\ref{sec:spectral-features-all})  & \numprint{\EBVIII}  \numprint{\eEBVIII}  & \numprint{\RVIII}  \numprint{\eRVIII}  & \numprint{\AVIII}  \numprint{\eAVIII} & \numprint{\DelmuIII}  \numprint{\eDelmuIII}
\enddata
\end{deluxetable}


\newpage
\subsection{Na I and DIB Features in SN 2012cu Spectra}\label{sec:ISM}

\ed{We will now examine the Na~I and two DIB absorption features in the SN~2012cu spectra.}  All three features, Na~I, DIB at 5780~\AA, and DIB at 6283~\AA, are at the same redshift as the H~I along the sightline toward \sncu in NGC~4772 \citep{haynes00a}.

\textbf{Na~I Absorption:} For known dust-to-gas ratio, the Na I column density can be used to infer the amount of dust along the line of sight.  In addition, the correlation between extinction and the EW of Na~I was first noticed decades ago \citep[e.g.][]{merrill38a} and has been extensively studied since \citep[e.g.][]{richmond94a, turatto03a, poznanski12a, phillips13a}. We can clearly see the Na~I D feature due to the doublet $\lambda\lambda$5896, 5890~\AA\ in the SN~2012cu spectra (Figure~\ref{fig:spect-comp}, \ref{fig:stacked-DIB5780}).  \citet{sternberg14a} obtained high resolution measurement of the neutral Na features of SN~2012cu at four epochs..  Their spectrum at a phase of 8 days gives (EW$_\mathrm{D1}$, EW$_\mathrm{D2}$) = ($857\pm6$~m\AA, $928 \pm 6$~m\AA), and shows that the lines are so saturated that the Na I column density cannot be measured accurately.  

We measure EW(NaD) in our low resolution SNIFS spectra for 16 epochs, spanning phases of \cuPzero to \cuPsixteen days, for which the S/N was adequate. Our model consisted of Gaussians for the D1 and D2 components, \edd{simultaneously fit with a background parametrized} by a fourth order polynomial.  The width of the Gaussians are set equal to \edd{the 3.42~\AA\ RMS of the SNIFS line spread function.}  The ratio of the lines was set to unity to reflect the strong saturation observed in the UVES spectrum. The \rf{SNIFS} spectrum taken at the same phase as the UVES spectrum yields EW(NaD)~\rf{$= 1790\pm80$~m\AA} 
in good agreement \rf{with the UVES results  $\mathrm{EW}_\mathrm{D1} + \mathrm{EW}_\mathrm{D2} = 1785 \pm8$~m\AA}.  \rf{For the full SNIFS spectral time series, we} find a mean \rf{$\langle \mathrm{EW(NaD)} \rangle = 1560$~m\AA} with \rf{$\mathrm{RMS} = 130$~m\AA}, with $\chi^2_{\nu}=3.0$. This \edd{EW} is marginally lower than the values from the UVES spectrum. But the somewhat high $\chi^2_{\nu}$ suggests the presence of extra noise, likely due to residual structure in SN spectral features that is difficult to completely remove from our low-resolution spectra. There is no indication of any coherent trend with phase. Because the Na~D line is so strongly saturated, only variability having a velocity at the extremities of the lines could have been detected.  

\ed{\textbf{DIB Features:} The DIBs have a well-known association with dust \citep[e.g.][]{herger22a, merrill34a, herbig95a, cox08a, hobbs08a}.  We find many DIB features in the high resolution VLT UVES spectrum of SN~2012cu at a phase of 8 days.  The DIBs at 5780~\AA\ and 6283~\AA\ are strong enough to be found in the SNIFS spectra (Figures~\ref{fig:spect-comp}, \ref{fig:stacked-DIB5780}).  \citet{phillips13a} used the EW of the DIB feature at 5780~\AA\ to measure extinction for a number of SNe Ia.  We find this feature in our spectra for phases \cuPzero, \cuPone, \cuPtwo, \cuPthree, and \cuPfour days (Figure~\ref{fig:stacked-DIB5780}).  This feature is located in a Si~II absorption line, making it challenging to fit.  \edd{The approach that works best} is to simultaneously fit this absorption feature and the background in a 100~\AA\ window around 5780~\AA\ with a Gaussian and a third order polynomial (Figure~\ref{fig:DIB-fit}).  The convolution of this feature, with an intrinsic FWHM of 2.11~\AA\ \citep{hobbs08a, phillips13a}, and the SNIFS line-spread-function for the red channel yields a net RMS width of 3.54~\AA.  We thus fix the width of the Gaussian at this value.  (From the UVES spectrum we also identify the weak 5800~\AA\ DIB, of which we can see a hint in the panel for phase \cuPone days in Figure~\ref{fig:DIB-fit}.  At the existing S/N, it does not affect our fit.)  After normalizing by the fitted the background, we find the values \EWDIB = 292$\pm$97, 390$\pm$76, 355$\pm$62, 370$\pm$74, and 432$\pm$71~m\AA\ for these \rf{five} phases, respectively.  
The weighted mean is 357$\pm$34~m\AA, and there is no evidence of time variation. 
From the high resolution VLT UVES spectrum at 8~days we measure \EWDIB~$= 360\pm5$~m\AA, in agreement with the results from the lower resolution SNIFS spectra.}

\noindent
\begin{minipage}{\linewidth}
\makebox[\linewidth]{
  \includegraphics[keepaspectratio=true,scale=0.4]{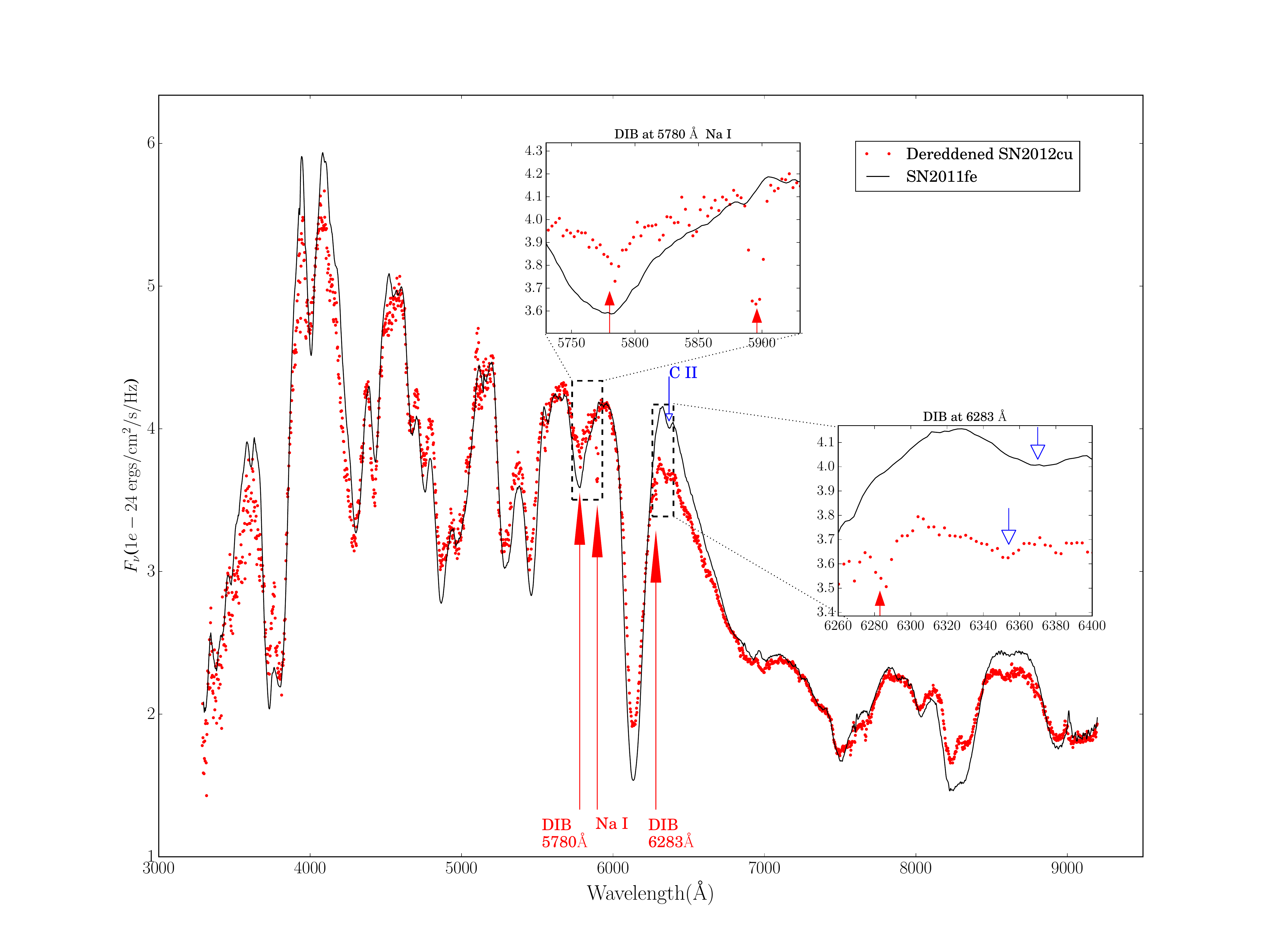}}
\captionof{figure}{Na~I and DIB features.  The solid line (black) and the dots (red) show the spectra of SN 2011fe at \fePtwo days and SN 2012cu at \cuPtwo days, respectively.  The spectrum of SN~2012cu has been dereddened according to the best-fit \EBV and \RV (see Figure~\ref{fig:all-phases-convl} or Table~\ref{tab:spect-diff-params}) for this phase.  Its flux has also been \edd{scaled} according to the best-fit \Delmu to correspond to the distance of SN~2011fe. The DIB absorption at 5780~\AA\ and the Na~I feature can be clearly seen (red arrows with filled arrowheads).  We also indicate another known DIB feature at 6283~\AA\ (red arrow with filled arrowhead).  All three features are at the redshift of the H~I toward SN2012cu in NGC~4772 \citep{haynes00a}.  The intrinsic C~II $\lambda6580$~\AA\ feature is also shown (blue arrow with unfilled arrowhead). 
}\label{fig:spect-comp} 
\end{minipage}

\noindent
\begin{minipage}{\linewidth}
\makebox[\linewidth]{
  \includegraphics[keepaspectratio=true,scale=0.5]{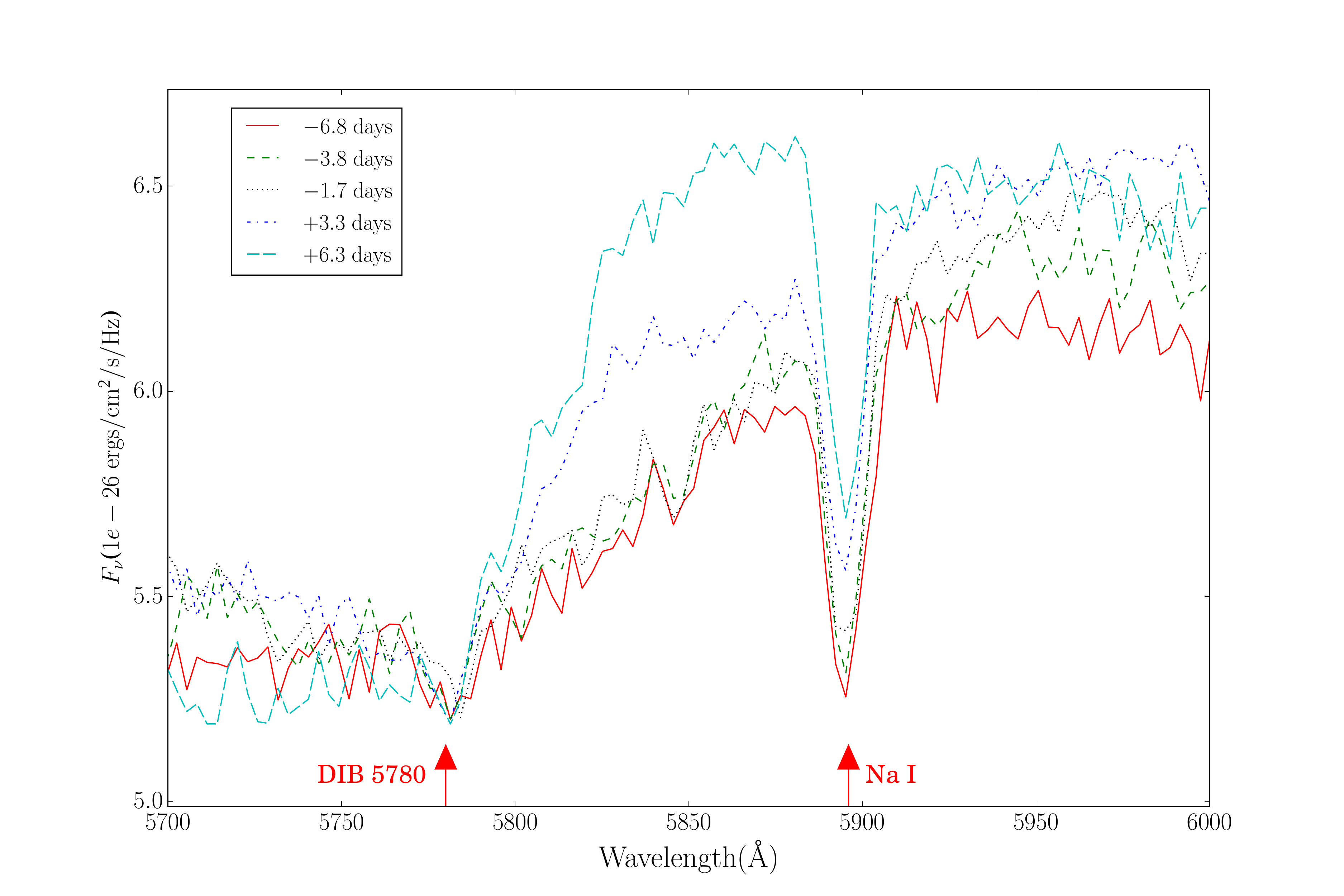}}
\captionof{figure}{Zoom around key ISM features.  \rf{The solid (red), dashed (green), dotted (black), dash-dotted (blue) and long-dashed (cyan) lines} are SN 2012cu spectra flux-matched at the bottom of a notch near 5780~\AA\ for the phases of \cuPzero, \cuPone, \cuPtwo,  \cuPthree, and \cuPfour days, respectively.  
The DIB absorption feature at 5780~\AA\edd{, situated near the minimum of a Si~II feature,} can now be clearly seen.  The Na~I absorption feature to the right is also readily apparent. 
}\label{fig:stacked-DIB5780}  
\end{minipage}

\noindent%
\begin{minipage}{\linewidth}
\makebox[\linewidth]{
  \includegraphics[keepaspectratio=true,scale=0.38]{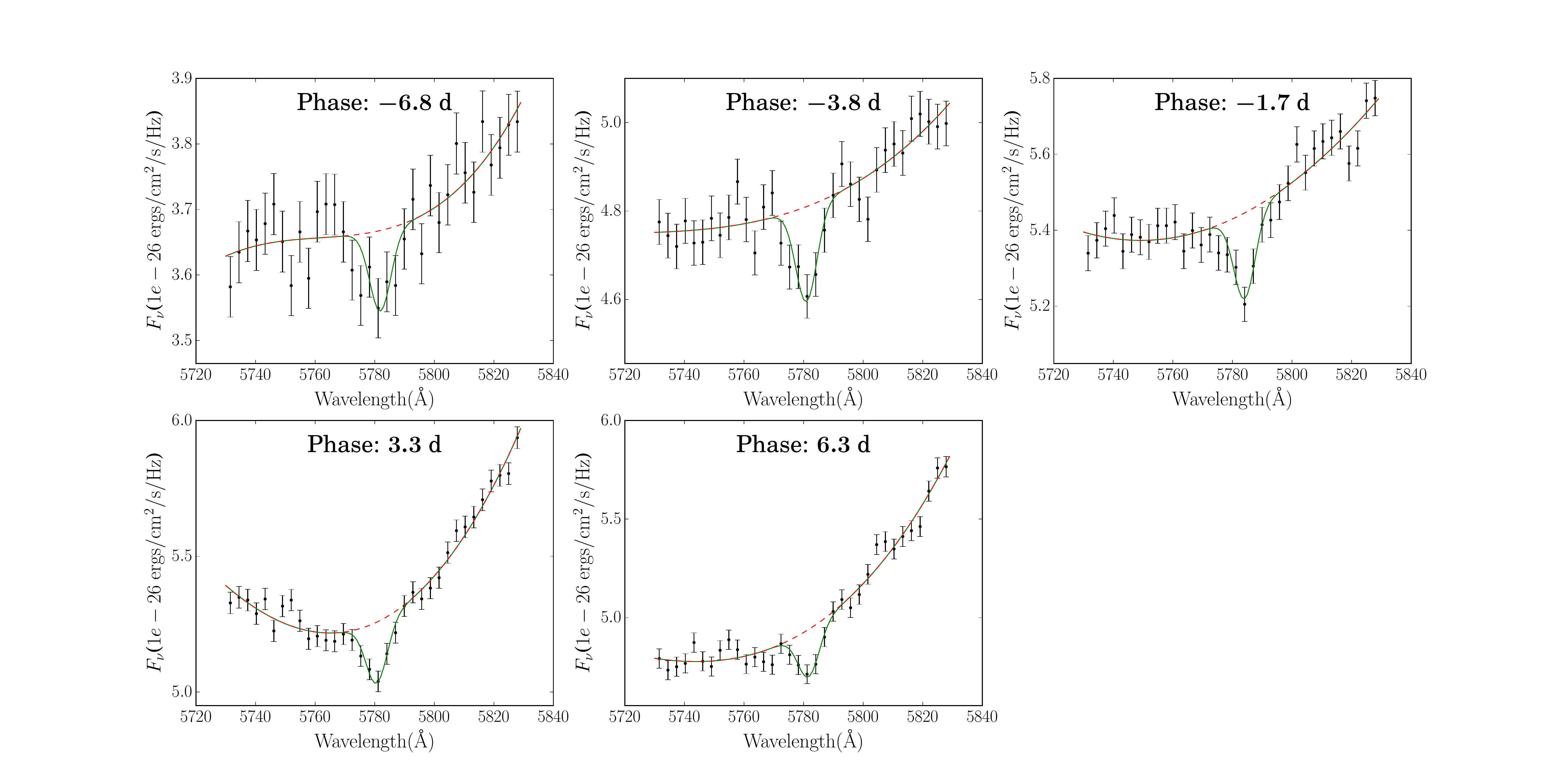}}
\captionof{figure}{\ed{Zoom of the spectral region covering the DIB 5780~\AA\ feature \edd{identified in Figure~\ref{fig:stacked-DIB5780}} for each epoch where it is well-detected.  The green curve in each panel shows a simultaneous fit with a Gaussian, and a third order polynomial for the background (red dashed curve).  We normalize by the background and determine the EW for each epoch.  We find an average EW of 357$\pm$34 m\AA.  This mean value is confirmed by our measurement of 360$\pm5$~m\AA\ for the EW of the same feature from a high resolution VLT UVES spectrum at a phase of 8 days.  This is about 1\,$\sigma$ below the MW average, 609$^{+402}_{-242}$ m\AA, expected using our best-fit \AV found in \S~\ref{sec:spectral-features-all} \citep[see][]{phillips13a}.}    
}\label{fig:DIB-fit} 
\end{minipage}

\newpage
\section{Discussion} \label{sec:discussion}

We now discuss some of the implications of our measurements, specifically, the distance to the SN~2012cu host galaxy and the nature of the veiling dust.
\subsection{Measurement of Host Distance}\label{sec:host-dist}

The distance modulus differences between SN 2012cu and SN 2011fe (\Delmu) from \ed{Approach~I (\S\ref{sec:straight-binned-fit}), II, and III (both in \S\ref{sec:spectral-features-all})} are in agreement with each other --- to better than their respective RMS dispersions (Table~\ref{tab:summary-3-approaches}).  \edd{As presented in \S~\ref{sec:spectral-features-all}}, the average value over 10 epochs from Approach II has been adopted as the best-fit \edd{relative distance modulus: (\Delmu, RMS) = (\numprint{\DelmuII}, \numprint{\eDelmuII})~mag}.  \ed{If we combine the gray scatter for SN~2012cu and SN~2011fe mentioned in \S~\ref{sec:straight-binned-fit}, we expect an RMS across the phases to be 0.052 mag.  With 10 phases, we expect the uncertainty on the RMS to be 0.013~mag.  Thus it is likely that the gray scatter can fully explain the RMS of \eDelmuII~mag for \Delmu.  But it is also possible that there are additional sources of uncertainty that may not be statistical.  Below we consider both possibilities.}

The distance to \ed{the host galaxy of SN 2011fe,} M101, has been measured most recently and accurately using the Cepheid period-luminosity relation \citep{shappee11a, mager13a, nataf15a}
and ``tip of the red giant branch'' \citep[TRGB;][]{shappee11a, lee13a} techniques. Adjusting these published values to a common distance modulus
of $18.477\pm0.033$~mag for the LMC \citep{freedman12a} gives a distance modulus to M101 of $29.114\pm0.072$~mag. The RMS scatter between
the five measurements is 0.10~mag 
and the weighted means of the Cepheid and TRGB measurements agree within 0.04~mag.  Adopting this distance to
M101 with our measured \Delmu results in a distance modulus of 31.11$\pm$0.15~mag for NGC~4772, the host of SN~2012cu.  This corresponds to 16.6$\pm$1.1 Mpc.
SNe~Ia exhibit a random per-object scatter of $\sim0.15$~mag when standardized using lightcurve width and color. However, closely matched ``twin'' SNe~Ia like SN~2011fe and SN~2012cu show a much smaller per-object scatter.  \ed{For the twinness ranking of the pair SN~2012cu and SN~2011fe (\edd{at top 10\%}), the scatter is expected be $0.086$~mag \citep{fakhouri15a}.  Removing the gray scatter of 0.025~mag quoted there, we obtain 0.082 mag.  If we treat the RMS dispersion for \Delmu as purely statistical, the uncertainty on \Delmu is $\sigma_{\Delta \mu} = \sqrt{2\cdot0.082^2 + \eDelmuII^2/(10 - 1)}$ = 0.12~mag.  If we take the extreme position that all of the RMS dispersion comes from non-statistical sources, then  $\sigma_{\Delta \mu}$ \edd{becomes} 0.13~mag.  Thus for the host galaxy of SN~2012cu, after adding the uncertainty for the SN~2011fe distance, $\sigma_{\mu}$ = 0.14 or 0.15 mag, for these two possibilities respectively.}  

NGC~4772 is usually considered part of the Virgo Southern Extension \citep{tully82a}.  \ed{Using an infall model, \citet{kim14a} concluded NGC~4772 is a member of the Virgo Cluster proper.}  A previous determination of its distance modulus can be found on the Extragalactic Distance Database\footnotemark[2] \citep[EDD, ][]{tully09a}: 30.96 $\pm$ 0.13 mag, corresponding to 15.6$\pm$1 Mpc.  This is in excellent agreement with our value.  Interestingly, the specific distance to NGC~4772 quoted in \citet{tully08a} is highly discrepant with both the EDD distance, \ed{which is the average distance for a group of 11 galaxies, \edd{to which NGC~4772 is assigned}}, and our value.  \rf{\citet{haynes00a} found that the stars and the ionized gas in the center of NGC~4772 are counter-rotating, which is often taken as an indication of a merger.}  If the bulk of the gas in NGC~4772 is from a merger, the H~I line-width would likely be inflated.  For the Tully-Fisher relation, this would result in a measured distance that is too large.  Therefore we consider the EDD value, rather than the value from \citet{tully08a} for this galaxy, as the most appropriate for this comparison.

\ed{If on the other hand, had we \edd{mimicked} a cosmological fit, e.g., using a value of $\beta\sim3.1$ \citep{betoule14a} and combining it with the SALT $c = 0.971$ for SN~2012cu, the estimated distance to NGC~4772 would have been 26 Mpc.  This would place NGC~4772 well beyond the distance of the Virgo Cluster.  Given the uncertainty surrounding the integrity of the H I line width for an apparent merger like NGC~4772, we advocate another independent measurement of the distance, e.g., using the TRGB or the surface brightness fluctuation (SBF) method.  If the TRGB or SBF distance also agrees with our distance measurement, that would further validate that our approach of correcting for the extinction using the twins method yields the correct distance.}

\footnotetext[2]{http://edd.ifa.hawaii.edu/}


\subsection{Nature of Dust}\label{sec:dust-nature}

\subsubsection{Upper Limit on Time Variability}\label{sec:lim-time-variable}

Figures~\ref{fig:params-phases-summary} shows that the dispersions for the dust quantities, \RV, \EBV, and \AV are small.  There is the suggestion of shallow trends in phase for \RV and \EBV, but in opposite directions.  \rf{The trends appear stronger for Approach I (Figure~\ref{fig:params-phases-summary}, left column).  The effects might be real, but consider the following:  1.} Systematic uncertainties unrelated to dust, e.g., slightly different SN astrophysics even for good but not perfect ``twins," likely remain.  \rf{The (likely small) difference in their time evolution may be responsible for the these trends.  The fact that when we de-weight the spectral feature differences in Approach II, the trends in \EBV and \RV clearly become even weaker is consistent with this conjecture.  2. The remaining very shallow trends in \EBV and \RV in Approach II (Figure~\ref{fig:params-phases-summary}, right column) still point in opposite directions.}  If there were time variation associated with dust, whether due to ISM or CSM, the first order effect would be the time variation of \AV, with \EBV following the same trend (e.g., for ISM, see \citet{patat10a}, \citet{foerster13a}; for CSM, see \citet{wang05a}, \citet{goobar08a}, \citet{brown15a}).  However, there is no clear trend for \AV in Figure~\ref{fig:params-phases-summary}.  While our model does not capture what may be responsible for the very weak trends that remain in \RV and \EBV in Approach II, their opposite directions may be related to the per-phase anti-correlation between these two quantities noted in \S~\ref{sec:straight-binned-fit}.  We therefore hesitate to claim that the apparent but shallow trends in phase for \RV or \EBV are real.  At this point we regard the RMS values of the fluctuations in \RV and \EBV as the upper limits for their time variation.  To confirm subtler variability would require a pair of SNe even more closely twinned than SN 2011fe and SN 2012cu.  \rf{Finally, while it is true that the systematic scatter is greater than the error bars from fitting (Figure~\ref{fig:params-phases-summary}), the spectral time series for these two SNe were measured so well that the fitting errors are far below the scale of interest scientifically.  The systematic scatters themselves are small.  Thus, even if these shallow trends were real, we expect the implications would be minor.}



\subsubsection{Lack of Evidence for CSM Dust}\label{sec:not-CSM}

\ed{Simulations of extinction due to CSM dust carried out by \citet{wang05a} and \citet{brown15a} show that for the phase range probed in this paper, \EBV and \AV should monotonically decrease over time with a similar fractional rate, and as a result \RV would only vary slightly.  Those simulated changes in \AV and \EBV are much greater than seen in Figure~\ref{fig:params-phases-summary}.  Changes in CSM ionization conditions due to SN ultraviolet radiation can also lead to variable EW(Na~I) \citep{patat07a}.  However, there is no evidence for EW(Na~I) variation in \citet{sternberg14a} nor in our measurements of the SNIFS spectra.}


\subsubsection{Evidence for ISM Dust}\label{sec:evidence-ISM}
\ed{The \RV value we have obtained, $\RVII \pm \eRVII$, is very similar to MW average.  The expected EW for DIB~5780~\AA\ based on MW observations from Equation~6 of \citep{phillips13a} is $609^{+406}_{-242}$~m\AA, and the direct measurement for NGC 4772, 357$\pm$34~m\AA, \edd{agree within the expected scatter}.  In addition, as mentioned in \S~\ref{sec:ISM}, we see many other DIB features in the VLT UVES spectrum.  Using the MW average relationship between EW(Na~I) and $E(B-V)$ in \citet{poznanski12a} and the total EW(Na~I)  from \citet{sternberg14a}, \citet{amanullah15a} obtained an \EBV prediction that is approximately 2\,$\sigma$ 
higher than the measured value for SN~2012cu.  \edd{To some degree this is expected given the very strong saturation of the Na~I~D absorption and is consistent with the larger scatter on \AV vs. EW(Na~I) found for SNe by \citet{phillips13a}.}} 

Lending support to ISM dust being the cause of the reddening for SN~2012cu is the projected position of SN~2012cu onto a dust lane in its host galaxy (Figure~\ref{fig:acq-image-WISE}).   
\rf{The disk rotation of NGC~4772 was measured by \citet{haynes00a} who identified an inner ring of H~I that is co-spatial with the dust ring.  At the projected location of \sncu, the H~I velocity contours of their Figure 12 give a mean of 90~km/s.
From the UVES spectrum, we have measured the velocities of the Na~I~D and DIB~5780~\AA\ extinction tracers to be 89~km/s and 102~km/s, respectively.  These values match well with the H~I velocity at the location of \sncu.  \citet{haynes00a} also provided the bulge-disk decomposition of stellar light, which at the location of \sncu has a bulge-to-disk flux ratio of $\sim 0.07$, indicating that \sncu is most likely in the disk of NGC~4772.  Taken together, it is therefore highly likely that \sncu is embedded in or behind the ISM dust lane.}

In summary, we find that the observational evidence strongly suggests that the dominant, if not all, foreground dust component in the host of SN~2012cu is interstellar in nature and is similar to MW dust.
 
\ed{An interesting new avenue for constraining $R_V$, independent of SN~Ia colors and astrophysics, is the correlation between the apparent dust emissivity power law index, $\beta_{FIR}$, and $R_V$, found for Milky Way sightlines by  \citet{schlafly16a}.  It so happens that NGC~4772 was observed in the thermal IR using {\it Herschel}, yielding $\beta_{FIR}=2.1\pm0.3$ \citep{cortese14a}.  \edd{While typical for many galaxies,} this is outside the range spanned by the MW \citet{schlafly16a} data.  But if a linear extrapolation is applicable, it would predict $R_V(\beta_{FIR}) = 2.2\pm0.7$.  This is lower than, but consistent with our measurement based on twin SN~Ia optical spectra. $\beta_{FIR}$ has well-established correlations with galaxy metallicity, gas mass fraction, and stellar mass surface density \citep{cortese14a}.  Unless these differences translate directly into differences in $R_V$ --- a situation that would be relevant for extinction correction of SNe~Ia --- then differences between the properties of the NGC~4772 dust ring and the Milky Way \edd{dust} could bias this method. The size of the uncertainty further illustrates that it may prove challenging to measure $\beta_{FIR}$ with sufficient accuracy to help resolve the tension between the cosmologically-derived $\beta$ and SN-color based measurements of $R_V$.}

It is worth noting that there is \ed{at least one other} SN Ia with high extinction and a MW-like \RV, SN 1986G, with $A_V\sim2.0$ and \RV = $2.57^{+0.23}_{-0.21}$  \citep{phillips13a} based on optical and NIR photometry\footnotemark[3].  An earlier \RV value of 2.4 for SN 1986G was reported by \citet{hough87a} based on polarimetry and the Serkowski \edd{relation between peak polarization wavelength and \RV} \citep{serkowski75a}.  Even though SN 1986G is considered a weak and peculiar event \citep{phillips87a, branch98a}, 
the host galaxies for SN 1986G and SN 2012cu, NGC 5128 and NGC 4772, respectively, are both large, mostly passive galaxies with a ``frosting" of young stars and dust lanes indicative of a merger \Cite{israel98a}{haynes00a}.  Like SN~2012cu, SN 1986G is projected onto its host galaxy dust lane and the \ed{dust lane is considered to be the source of its reddening \citep{phillips87a, phillips13a}.} 

\footnotetext[3]{Even though \citet{phillips13a} used the extinction curve of CCM89, their \RV value for SN~1986G might not be severely underestimated due to \edd{their inclusion of a larger wavelength coverage and the dilution of the 5500 -- 8900~\AA\ region by the addition of NIR photometry (see Appendix~\ref{sec:appendixA}).}}

\section{Conclusions} \label{sec:conclude}
We have used a high-quality spectrophometric time series obtained by the SNfactory to study the highly reddened Type~Ia SN~2012cu.  We analyzed \edd{10 phases between \cuPzero and \cuPeleven~days}, using the phase-matched spectra of SN 2011fe as an unreddened template.  By simultaneously fitting for \EBV, \RV, and \Delmu, the distance modulus difference between the two SNe, we have found that SN 2012cu is highly reddened, with (\EBV, RMS)~$ = (\numprint{\EBVII},  \numprint{\eEBVII})$, (\RV, RMS)~$= (\numprint{\RVII},\numprint{\eRVII})$,  (\AV, RMS)~$ = (\numprint{\AVII}, \numprint{\eAVII})$, and (\Delmu, RMS)~$= (\numprint{\DelmuII},  \numprint{\eDelmuII}$).  Our best-fit \EBV agrees well with the corresponding value reported in \citet{amanullah15a} for SN 2012cu based on broadband photometry from UV to NIR, and our best-fit \RV is slightly higher than theirs.
 
 {\bf Spectral Diversity and Relative Distance Measurement with Supernova Twins}.  
 We find that \sncu and \snfe are excellent spectroscopic twins according to the method of \citet{fakhouri15a}.  We have treated the modest spectral differences that remain in different ways.  The consistency of our measurement of \AV and \Delmu between these approaches demonstrates that despite the presence of spectral feature differences, spectroscopically twinned SNe can be used to accurately measure relative distances, in this case between M101 and NGC 4772, the host galaxies to the two SNe, to 6.0\%.  
 
{\bf Host Distance}.  We measure the distance modulus to the host galaxy of SN 2012cu, NGC 4772, to be $31.10\pm0.15$ mag, corresponding to a distance of 16.6$\pm$1.1 Mpc.  Our result is in excellent agreement with the value found on EDD. 

{\bf Nature of Dust Toward SN 2012cu}. \edd{While it is difficult to completely eliminate a contribution from circumstellar dust, together the following factors strongly suggest that the dominant dust component along the line of sight to SN 2012cu is interstellar in nature and is similar to MW dust:}

\rf{1) it's highly likely that SN 2012cu is embedded in or behind its host-galaxy dust lane;} 

\edd{2) the agreement between our best-fit \RV and the MW average value;}

 \edd{3) the lack of time variation in \EBV and \AV;}

\edd{4) the lack of time varying Na~I (or DIB 5780~\AA) absorption;}

\edd{5) our measurement of the 5780 \AA\ DIB and \AV agree with the average MW relation;}

\edd{6) the presence of many other DIB features identified in the high resolution VLT UVES spectrum;}

\edd{7) our measurement of Na I absorption and \AV agree with the broad trend seen in the MW.}


\newpage
\section{Acknowledgement} \label{sec:acknowledge}
We thank the technical staff of the University of Hawaii 2.2 m
telescope, and Dan Birchall for observing assistance.
We recognize the significant cultural role of Mauna Kea
within the indigenous Hawaiian community, and we appreciate the
opportunity to conduct observations from this revered site.  
This work was supported in part by the Director, Office of Science,
Office of High Energy Physics of the U.S. Department of Energy
under Contract No. DE-AC02- 05CH11231.  \rf{We thank the Gordon \& Betty Moore Foundation for their
continuing support.}
XH acknowledges the University of San Francisco (USF) Faculty Development Fund.  
ZR was supported in part by a USF Summer Undergraduate Research Fellowship. 
Support in France was
provided by CNRS/IN2P3, CNRS/INSU, and PNC; LPNHE acknowledges
support from LABEX ILP, supported by French state funds managed by
the ANR within the Investissements d'Avenir programme under reference
ANR-11-IDEX-0004-02.  NC is grateful to the LABEX Lyon Institute
of Origins (ANR-10-LABX-0066) of the Universit\'e de Lyon for its
financial support within the program ``Investissements d'Avenir''
(ANR-11-IDEX-0007) of the French government operated by the National
Research Agency (ANR).  Support in Germany was provided by the DFG
through TRR33 ``The Dark Universe;'' and in China from Tsinghua
University 985 grant and NSFC grant No~11173017.  \rf{This works is partially based on 
observations made with ESO Telescopes at the La Silla Paranal Observatory, Chile, under program 289.D-5035.}
Some results were
obtained using resources and support from the National Energy
Research Scientific Computing Center, supported by the Director,
Office of Science, Office of Advanced Scientific Computing Research of the U.S. Department of Energy
under Contract No. DE-AC02- 05CH11231.

\bibliography{dustarchive}

\newpage
\appendix

\section{Comparison of Three Extinction Curve Models} \label{sec:appendixA}

\ed{Here we compare the extinction curve models of CCM89, OD94, and F99 \edd{for our application}.  We choose the SN~2012cu phase of \cuPsix~days (pair \pound6 in Table~\ref{tab:phase-diff-meas-error-params}), for which the spectral features of SN~2011fe and SN~2012cu are the most similar.  Examination of several phases indicates that the conclusions presented below are independent of the phase chosen.}  To provide an initial assessment of these issues, we first use the best-fit \EBV, \RV, and \Delmu found for F99 in \S~\ref{sec:straight-binned-fit} to deredden SN 2012cu with each of these three extinction curves and compare to the \snfe spectrum at \fePsix~days.  We present the results in Figure~\ref{fig:redlaw_comp}.  For OD94 and CCM89, while the fits are virtually indistinguishable from F99 \edd{for wavelengths} blue-ward of 5200 \AA, there clearly is tension \edd{for these two curves} with the data in the wavelength region of 5500 -- 8900 \AA.  This region roughly corresponds to the $r$ and $i$ bands, \ed{and the tension between extinction curves here was first pointed out by F99.  For MW sight lines, when \RV~$ = 3.1$ is used for CCM89 and F99, \edd{they show that} CCM89 overcorrects in the wavelength range that corresponds to $r$ and $i$ bands (their Figure 6).} 

Figure~\ref{fig:redlaw_comp2} shows that this problem with CCM89 and OD94 persists even when \EBV, \RV, and \Delmu are optimized for each extinction curve model individually.  For this comparison we used Approach I from \S\ref{sec:straight-binned-fit}, where only measurement uncertainties are taken into account, since the de-weighting in Approach II has the potential to suppress regions that differ due to the extinction curve rather than SN features.  In this case, the best-fit \RV values are (CCM89, OD94, F99) = (2.23, 2.43, 2.99), with corresponding $\chi^2_\nu$ = (27, 28, 20).   While OD94 and CCM89 now match the data more closely in the wavelength region of 5500 -- 8900 \AA, clearly their agreement remains inferior to F99.  Furthermore, for OD94 and CCM89 the fit in the range 5000 -- 5500 \AA\ has worsened in a systematic way.  \ed{That the \RV values for CCM89 and OD94 are now lower is not a surprise: given the tension seen in Figure~\ref{fig:redlaw_comp}, CCM89 and OD94 need to be ``steeper" to fit the the wavelength range of 5500 -- 8900 \AA\ better, which translates to a lower best-fit \RV.}

In both comparison approaches, F99 clearly provides the best fit, and OD94 the worst.  This finding is in agreement with \citet{schlafly10a}, who compared the predictions of the same three extinction curve models with the colors of stars from the Sloan Digital Sky Survey (SDSS) and 
found that F99 fit the data the best, and OD94 the poorest.  The extinction curves of OD94 and, to a lesser extent, of CCM89, were disfavored because they under-predicted the difference between the $r-i$ and $i-z$ colors (see Figure 18 of \citet{schlafly10a}).  These correspond to the same wavelength regions where we have found tension for OD94 and CCM89.  \citet{berry12a} compared the predictions of the same three extinction curves with stellar photometry from SDSS, and they also found OD94 to be inconsistent with data, although they considered CCM89 to be acceptable.

\noindent
\begin{minipage}{\linewidth}
\makebox[\linewidth]{
  \includegraphics[keepaspectratio=true,scale=0.4]{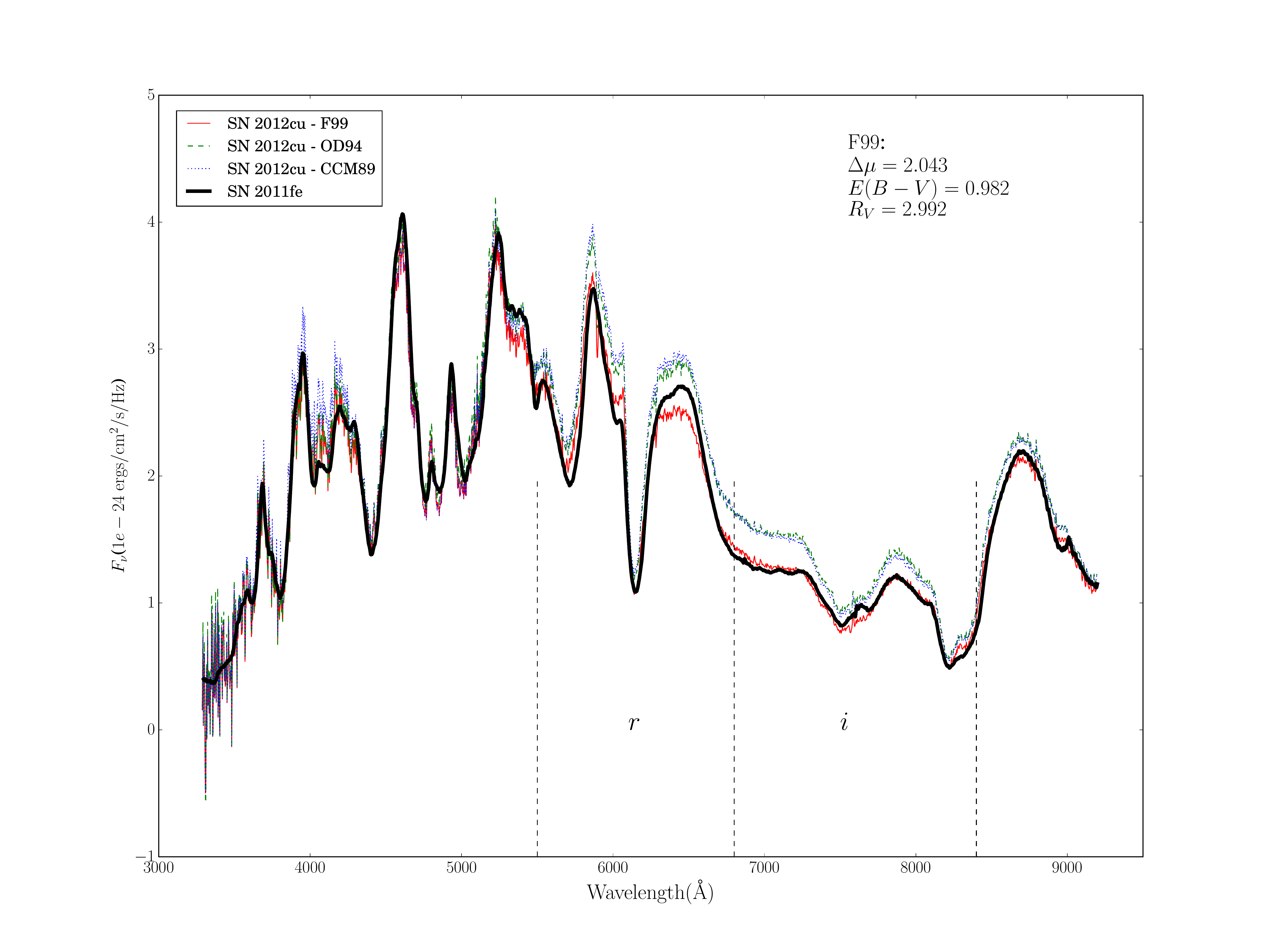}}
\captionof{figure}{Using the best-fit \EBV, \RV, and \Delmu for the F99 extinction curve model by Approach I in \S~\ref{sec:straight-binned-fit} (see Table~\ref{tab:spect-diff-params}), the spectrum of SN 2012cu at \cuPsix~days is dereddened with the F99 (red line), OD94 (green dashed line), and CCM89 extinction curve models (blue dotted line), and then adjusted by \Delmu to match the brightness of SN~2011fe.  Also plotted is the spectrum of SN~2011fe at \fePsix~days
(thick black line), used as the template.  The vertical dashed lines demarcate, roughly, the boundaries of the $r$ and $i$ filters.  
}\label{fig:redlaw_comp}
\end{minipage}

\noindent
\begin{minipage}{\linewidth}
\makebox[\linewidth]{
  \includegraphics[keepaspectratio=true,scale=0.4]{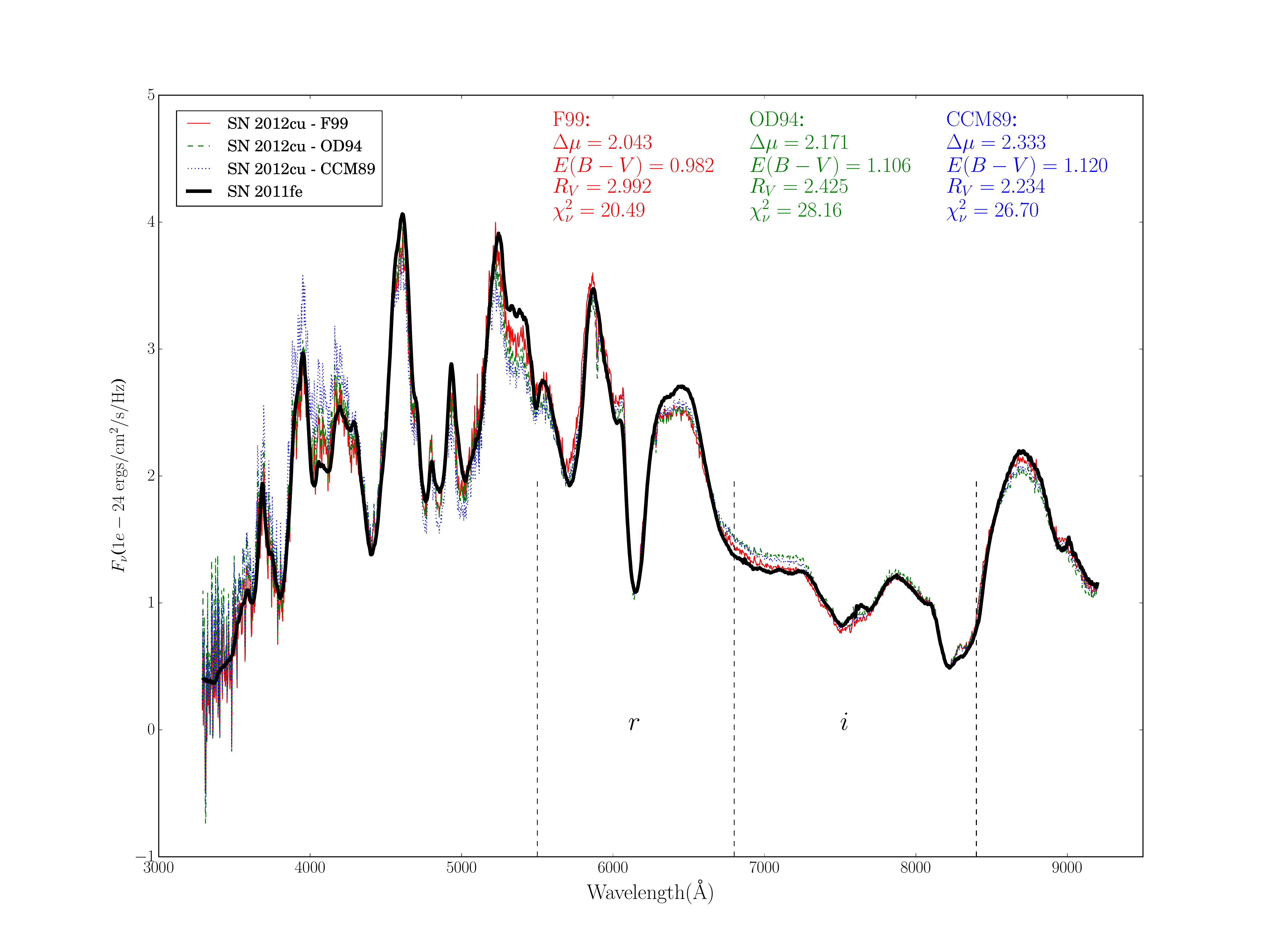}}
\captionof{figure}{Comparison of the best-fit \EBV, \RV, and \Delmu found for each of the three extinction curve models, using Approach I in \S\ref{sec:straight-binned-fit}.  The spectrum of SN 2012cu at \cuPsix~days is dereddened with F99 (red line), OD94 (green dashed line), and CCM89 (blue dotted line), adjusted according to \Delmu to match the brightness of SN 2011fe.  Also plotted is the spectrum of SN~2011fe at \fePsix~days
(thick black line), used as the template.  The vertical dashed lines indicate the boundaries of the $r$ and $i$ filters.  
}\label{fig:redlaw_comp2}  
\end{minipage}

In addition, using extragalactic sources to constrain the MW dust properties, \citet{mortsell13a} compared F99 and CCM89.  Though their analysis was based on colors, it is similar to ours in the sense that both \EBV and \RV were allowed to vary.  They found that the \RV values for CCM89 were consistently lower than those for F99.  For QSO's, brightest central galaxies, and luminous red galaxies, the \RV values were (CCM89, F99) = ($3.10\pm0.43$, $3.25\pm0.37$), ($2.89\pm0.88$, $3.42\pm0.48$), and ($1.95\pm0.35$, $3.16\pm 0.36$), respectively, where the uncertainties correspond to their 95.4\% confidence level.  Our results, as presented in Figure~\ref{fig:redlaw_comp2}, show the same pattern.  They also made a direct comparison between the extinction values of CCM89 and F99 for the same \RV.  From their Figure~13, it is clear that for \RV = 3, the extinction correction ($A_\lambda$) is less for F99 than for CCM89 in the wavelength region of the $r$ and $i$ bands.  This is exactly as F99 pointed out and as we have shown in our Figure~\ref{fig:redlaw_comp}.

\citet{amanullah15a} reported that for \sncu their best-fit \RV for F99 ($2.8\pm0.1$) and OD94 ($2.8\pm0.2$) are in agreement.  This may be due to the fact that the wavelength coverage in their analysis was from UV to NIR.  The wide wavelength coverage may have diluted 
the discrepancy in the wavelength range between 5500 \AA\ -- 8900\AA.  However, the uncertainty for their OD94 best-fit \RV is twice as large as that for F99.  We suspect that the larger uncertainty, and higher $\chi^2_\nu$, are due to the discrepancy between extinction curve models in the $r$ and $i$ wavelength range.

\newpage
\section{SN~2012cu Spectral Time Series Data} \label{sec:appendixB}
\setcounter{figure}{0}
\ed{Observations of SN~2012cu were obtained by the SNfactory with its SuperNova Integral Field Spectrograph (SNIFS).  SNIFS is a fully integrated instrument optimized for automated observation of point sources on a structured background over the full ground-based optical window at moderate spectral resolution.  It consists of a high-throughput wide-band pure-lenslet integral field spectrograph \citep[IFS;][]{bacon95a, bacon01a}, a parallel photometric channel to image the stars in the vicinity of the IFS field-of-view to monitor atmospheric transmission during spectroscopic exposures, and an acquisition/guiding channel.  The IFS possesses a fully-filled $6.4{\twopr} \times 6.4{\twopr}$ spectroscopic field of view subdivided into a grid of $15\twopr \times 15\twopr$ spatial elements, a dual-channel spectrograph covering 3200--5200~\AA\ and 5100--10000~\AA\ simultaneously, and an internal calibration unit (continuum and arc lamps). SNIFS is continuously mounted on the south bent Cassegrain port of the University of Hawaii 2.2~m telescope on Mauna Kea. The telescope and instrument, under script control, are operated remotely.}

\ed{SNfactory follow-up observations of SN~2012cu commenced on 2012, June 20.3, and span 53 nights. 
The observing log is presented in Table~\ref{tab:obs-log}.  The nominal observational cadence of 2 -- 3 nights was maintained until 31.4 days after maximum brightness.  The SN was observed for two more nights.  The last spectra reported here were obtained 2012, Aug. 12.3.  All spectra were reduced using SNfactory's dedicated data reduction pipeline \citep{bacon01a, aldering06a, scalzo10a}.  For non-photometric nights, we correct for clouds using field stars in the SNIFS parallel imaging camera.  Detailed discussions of the flux calibration and host-galaxy subtraction are provided in \citet{buton13a} and \citet{bongard11a}, respectively.  The final spectrophotometric time series of SN~2012cu in the observer's frame is displayed in Figure~\ref{fig:spectral-time}.  A tar file containing these spectra is available at http://snfactory.lbl.gov/TBD.}

\newpage
\begin{deluxetable}{cccccccc}
\tablecaption{\label{tab:obs-log} Observation Log}
\tablehead{\colhead{Phase (days)} & \colhead{UTC Date} & \colhead{MJD} & \colhead{Photometric} & \colhead{Standard Stars$^a$} & \colhead{Exp. Time(s)} & \colhead{Airmass} & \colhead{Seeing($^{{\prime\prime}}$)}}
\startdata
$-6.8$ & 2012 Jun. 20.3 & 56098.3 & N & 14 & 820.0 & 1.26 & 0.81 \\
$-3.8$ & 2012 Jun. 23.3 & 56101.3 & Y & 10 & 920.0 & 1.16 & 1.00 \\
$-1.7$ & 2012 Jun. 25.4 & 56103.4 & Y & 11 & 1620.0 & 1.81 & 1.82 \\
$3.3$ & 2012 Jun. 30.3 & 56108.3 & Y & 12 & 2440.0 & 1.65 & 1.65 \\
$6.3$ & 2012 Jul. 3.3 & 56111.3 & Y & 8 & 1220.0 & 1.62 & 1.03 \\
$8.3$ & 2012 Jul. 5.3 & 56113.3 & Y & 13 & 920.0 & 1.64 & 1.95 \\
$11.2$ & 2012 Jul. 8.3 & 56116.3 & Y & 19 & 1020.0 & 1.44 & 2.06 \\
$14.2$ & 2012 Jul. 11.3 & 56119.3 & N & 12 & 1020.0 & 1.53 & 1.09 \\
$16.2$ & 2012 Jul. 13.3 & 56121.3 & N & 10 & 920.0 & 1.69 & 1.31 \\
$18.2$ & 2012 Jul. 15.3 & 56123.3 & Y & 10 & 1520.0 & 1.74 & 1.27 \\
$21.2$ & 2012 Jul. 18.3 & 56126.3 & Y & 11 & 2740.0 & 1.66 & 2.16 \\
$23.2$ & 2012 Jul. 20.3 & 56128.3 & Y & 8 & 920.0 & 1.76 & 1.31 \\
$26.2$ & 2012 Jul. 23.3 & 56131.3 & N & 13 & 1520.0 & 1.62 & 1.31 \\
$28.2$ & 2012 Jul. 25.3 & 56133.3 & Y & 11 & 320.0 & 1.50 & 1.17 \\
$31.2$ & 2012 Jul. 28.3 & 56136.3 & Y & 5 & 1220.0 & 1.75 & 1.23 \\
$41.2$ & 2012 Aug. 7.3 & 56146.3 & Y & 9 & 820.0 & 2.07 & 1.37 \\
$46.2$ & 2012 Aug. 12.3 & 56151.3 & Y & 12 & 820.0 & 2.33 & 1.46 \\
\enddata
\tablenotetext{a}{Number of standard stars observed during the night and used for atmospheric extinction and telluric absorption correction.}
\end{deluxetable}

\newpage
\begin{minipage}{\linewidth}
\makebox[\linewidth]{
  \includegraphics[keepaspectratio=true,scale=0.47]{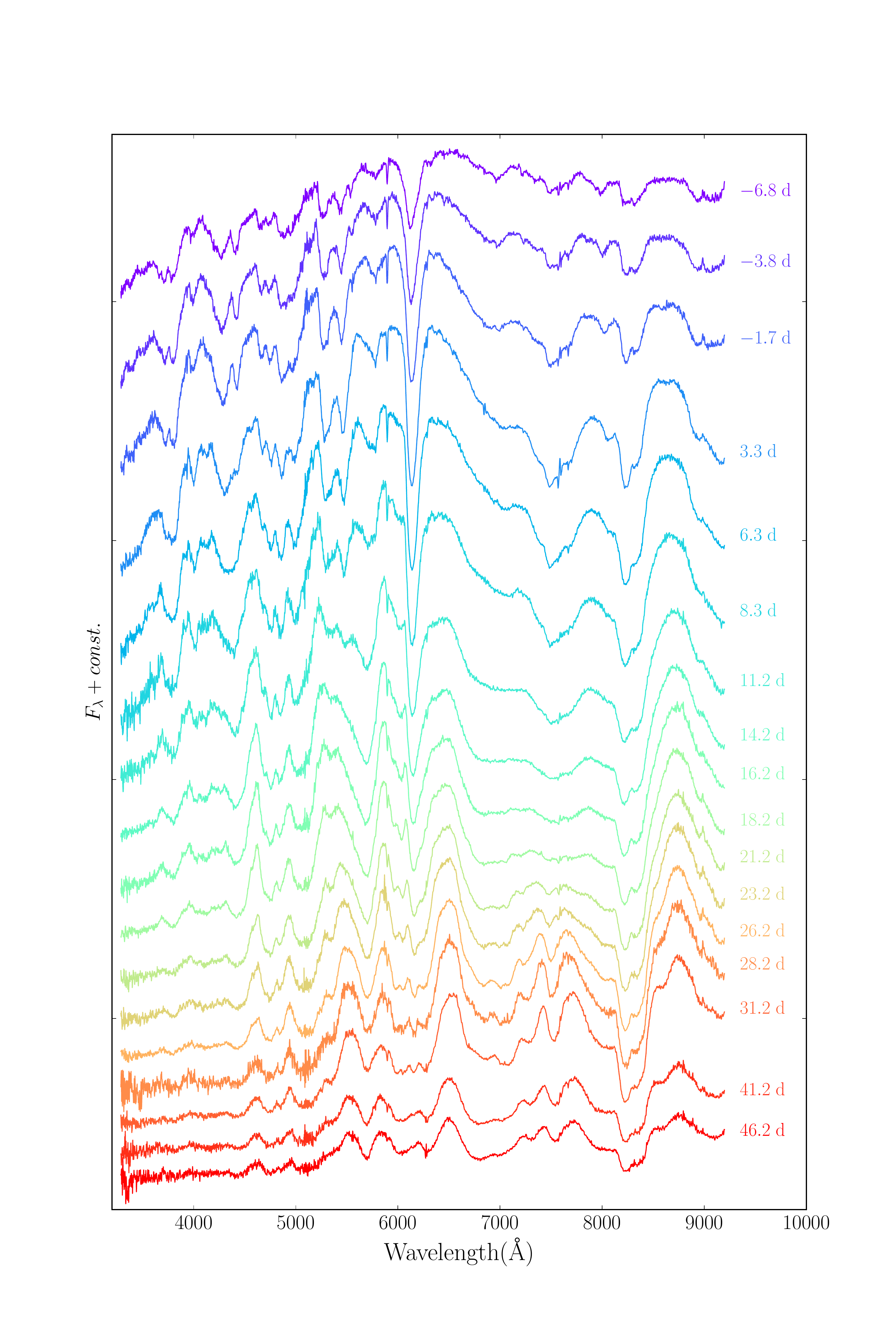}}
\captionof{figure}{SNIFS spectrophotometric time series of \sncu for phases of \cuPzero  to \cuPsixteen days. 
}\label{fig:spectral-time}
\end{minipage}


\end{document}